\documentclass[times,twocolumn,final]{elsarticle}

\usepackage{medima}
\usepackage{framed,multirow}
\usepackage{adjustbox}
\usepackage{amssymb}
\usepackage{latexsym}

\usepackage[table,x11names]{xcolor}
\usepackage{url}
\usepackage{xcolor}
\usepackage{amsmath}
\usepackage{hyperref}
\usepackage{amsmath}
\usepackage{amssymb}
\usepackage{dsfont}
\newcommand{\app}{\raise.17ex\hbox{$\scriptstyle\sim$}}
\usepackage{multirow}
\usepackage{lipsum}
\usepackage{times}
\usepackage{epsfig}
\usepackage{graphicx}
\usepackage{amssymb}
\usepackage{pifont}
\usepackage{hyperref}
\usepackage{soul}
\soulregister\cite7
\soulregister\ref7
\soulregister\text7

\definecolor{newcolor}{rgb}{.8,.349,.1}

\journal{Medical Image Analysis}

\begin{document}

\verso{Given-name Surname \textit{et~al.}}

\begin{frontmatter}


\title{UNesT: Local Spatial Representation Learning with Hierarchical Transformer for Efficient Medical Segmentation}

\author[1]{Xin \snm{Yu}}

\author[1]{Qi \snm{Yang}}
\author[1]{Yinchi \snm{Zhou}}


\author[2]{Leon Y. \snm{Cai}}

\author[1,3]{Riqiang \snm{Gao}}
\author[1]{Ho Hin \snm{Lee}}

\author[2]{Thomas \snm{Li}}

\author[4]{Shunxing \snm{Bao}}

\author[3]{Zhoubing \snm{Xu}}


\author[5]{Thomas A. \snm{Lasko}}

\author[2,6]{Richard G. \snm{Abramson}}

\author[7]{Zizhao \snm{Zhang}}
\author[1,4]{Yuankai \snm{Huo}}
\author[1,2,4,5]{Bennett A. \snm{Landman}}
\author[4,8]{Yucheng \snm{Tang}}

\address[1]{Department of Computer Science, Vanderbilt University, Nashville, TN, USA 37212}
\address[2]{Department of Biomedical Engineering, Vanderbilt University, Nashville, TN, USA, 37212}
\address[3]{Digital Technology and Innovation, Siemens Healthineers, Princeton, NJ, USA, 08540}
\address[4]{Department of Electrical and Computer Engineering, Vanderbilt University, Nashville, TN, USA, 37212}
\address[5]{Department of Biomedical Informatics, Vanderbilt University Medical Center, Nashville, TN, USA, 37235}
\address[6]{Annalise-AI, Pty, Ltd}
\address[7]{Google Cloud AI}
\address[8]{Nvidia Corporation}


\begin{abstract}

Transformer-based models, capable of learning better global dependencies, have recently demonstrated exceptional representation learning capabilities in computer vision and medical image analysis. Transformer reformats the image into separate patches and realizes global communication via the self-attention mechanism. However, positional information between patches is hard to preserve in such 1D sequences, and loss of it can lead to sub-optimal performance when dealing with large amounts of heterogeneous tissues of various sizes in 3D medical image segmentation. Additionally, current methods are not robust and efficient for heavy-duty medical segmentation tasks such as predicting a large number of tissue classes or modeling globally inter-connected tissue structures. To address such challenges and inspired by the nested hierarchical structures in vision transformer, we proposed a novel 3D medical image segmentation method (UNesT), employing a simplified and faster-converging transformer encoder design that achieves local communication among spatially adjacent patch sequences by aggregating them hierarchically.
We extensively validate our method on multiple challenging datasets, consisting of multiple modalities, anatomies, and a wide range of tissue classes, including 133 structures in the brain, 14 organs in the abdomen, 4 hierarchical components in the kidneys, inter-connected kidney tumors and brain tumors. We show that UNesT consistently achieves state-of-the-art performance and evaluate its generalizability and data efficiency. Particularly, the model achieves whole brain segmentation task complete ROI with 133 tissue classes in a single network, outperforming prior state-of-the-art method SLANT27 ensembled with 27 networks. Our model performance increases the mean DSC score of the publicly available Colin and CANDI dataset from 0.7264 to 0.7444 and from 0.6968 to 0.7025, respectively. Code, pre-trained models, and use case pipeline are available at: \url{https://github.com/MASILab/UNesT}.

\end{abstract}


\end{frontmatter}

\section{Introduction}
Medical image segmentation tasks have become increasingly challenging due to the need for modeling hundreds of tissues~\citep{huo20193d,wasserthal2022totalsegmentator} or hierarchically inter-connected structures~\cite{landman2015miccai} in 3D volumes. 
In the past few years, convolutional neural networks (CNNs) have dominated medical image segmentation due to their superior performance. Among all the CNNs, the U-Net \cite{ronneberger2015u} and its variants have been the most widely used for medical image segmentation. A "U-shape" model generally consists of an encoder for global representation learning and a decoder to gradually decode the learned representation to a pixel-wise segmentation. However, CNN-based models' encoding performance is limited because of their localized receptive fields \cite{hu2019local}.

Vision Transformers (ViT), on the other hand, are capable of learning long-range dependencies and have recently demonstrated exceptional representational learning capabilities and effectiveness in computer vision and medical image applications \citep{dosovitskiy2020image,hatamizadeh2022unetr,zhou2021nnformer}. Unlike CNNs, ViTs learn better long-range information by tokenizing images into 1D sequences and leveraging the self-attention blocks to facilitate global communication~\cite{hatamizadeh2022unetr}, which makes transformers better encoders. However, by tokenizing the image into 1D patches, transformers are less able to capture local positional information compared to CNNs, due to the lack of locality inductive bias inherent to CNNs~\citep{cordonnier2019relationship,dosovitskiy2020image}. To overcome this, ViT usually requires a large amount of training data which is expensive to acquire~\citep{tang2022self,zhou2021nnformer}. With small datasets in the medical field, insufficient data can lead to model inefficiency, especially when dealing with a large number of tissues of various sizes. Moreover, the self-attention mechanism for modeling multi-scale features for high-resolution medical volumes is computationally expensive~\citep{beltagy2020longformer,han2021transformer,liu2021swin}.

To improve representation learning in transformers in small datasets, recent works envision the use of local self-attention~\citep{liu2021swin,cao2021swin,han2021transformer}. To leverage information across embedded sequences, "shifted window" transformers~\cite{liu2021swin} have been proposed for dense predictions and modeling multi-scale features. However, these attempts aiming to adapt the self-attention mechanism by modifying patch communication often yield high computational complexity. In addition, the Swin transformer under-performs when datasets are small, or there are a large number of structures~\cite{liu2021swin}. 

Considering the advantages of hierarchical models~\citep{cciccek20163d,roth2018multi,tang2022self} on modeling heterogeneous high-resolution radiographic images and inspired by the aggregation function in the nested ViT~\cite{zhang2022nested}, we propose a Hierarchical hybrid 3D U-shape medical segmentation model with Nested Transformers (UNesT). Specifically, with nested transformers as the encoder, UNesT hierarchically encodes features with the 3D block aggregation function and merges with the convolutional-based decoder via skip connections at various resolutions to enable learning of local behaviors for small structures or small datasets. The 3D nested structure retains the original global self-attention mechanism and achieves information communication across patches by stacking transformer encoders hierarchically.

We perform extensive experiments to validate the performance of UNesT on the challenging whole brain segmentation task with 133 classes using T1 weighted (T1w) MRI images and a collected renal substructures 3D CT volumetric dataset with 116 patients on characterizing multiple kidney components including renal cortex, medulla and pelvicalyceal system with kidney function. We further evaluate UNesT on three widely-used public datasets Beyond The Cranial Vault (BTCV) \cite{landman2015miccai}, KiTS19 \cite{heller2021state}, and BraTS21 \cite{baid2021rsna} to illustrate the generalizability of UNesT. We compare UNesT to recent convolutional and transformer-based 3D medical segmentations baselines and conduct scalability and data efficiency analysis in a low-data regime. 

Our contributions to this work can be summarized as:
\begin{itemize}
\item[\textbullet] We introduce a novel 3D hierarchical block aggregation module, and propose a new transformer-based 3D medical segmentation model, dubbed UNesT. The model provides local spatial patch communication to better capture various tissues. This method achieves hierarchical modeling of high-resolution medical images and outperforms local self-attention variants with a simplified design compared to the "shifted window" module leading to improved data efficiency.

\item[\textbullet] We validate UNesT on a whole brain segmentation task that contains hundreds of classes. UNesT outperforms the current convolutional- and transformer-based single model methods. Our single model also outperforms the prior top method SLANT27 \cite{huo20193d} ensembled with 27 networks, and achieves new state-of-the-art performance.

\item[\textbullet] We collect and manually delineate the first in-house renal substructure dataset (116 CT subjects). We show that our method achieves state-of-the-art performance for accurately measuring cortical, medullary, and pelvicalyceal system volumes. We demonstrate the clinical utility of this work through accurate volumetric analysis, strong correlations, and robust reproducibility. We also introduce MONAI Bundle, a new plug-and-use framework for deploying models. Our codes, trained models, and tutorials are released for public availability. 

\item[\textbullet] We investigate model scalability and data efficiency in low-data regimes as well as the impact of the size of pre-training dataset.  We show the proposed method's generalizability by validating it on public datasets: BTCV, KiTS19, and BraTS2021.

\end{itemize}

\section{Related Works}
\textbf{Medical Segmentation with Transformers.} 
Transformer models demonstrate the ability of modeling longer-range dependencies for high dimension and high-resolution medical images in 3D Space. The scalability, generalizability, and efficiencies of ViT and hierarchical transformers enable stronger representation learning for dense predictions (e.g., pixel-to-pixel segmentation). Medical image segmentation tasks embed learning problems with multi-scale features instead of fixed scale, such as word tokens. To employ the vanilla Transformer~\cite{dosovitskiy2020image} for medical images, recent works proposed variant architectures that use ViT as network components. 

Transformer is known for its capability of capturing long-range dependencies but lacks inductive bias, which is inherent in CNNs. To reap the advantages of both CNNs and transformers, many efforts have been made to integrate the benefits of CNNs and transformers into a hybrid network. In the medical image segmentation domain, these works can be classified into three types: Transformer as main encoder, transformer as secondary encoder, and fusion model of both transformer encoder and CNNs encoder \cite{li2023transforming}.

When utilizing the Transformer as the primary encoder, the segmentation model usually includes a sequence of successive transformer blocks as the encoder. Various studies have used this design, such as UNETR, VT-UNet, and SwinUNETR \citep{hatamizadeh2022unetr, peiris2021volumetric, tang2022self}. The advantage of sequence-to-sequence modeling as the first embedding for medical images is to directly generate tokenized patches for the feature representation. Most of these methods connect a convolutional neural network (CNN)-based decoder and form the ”U-shaped” architecture for segmentation. This design features the long-range modeling ability for input images with a transformer encoder and better inductive bias with CNN decoder.

The second design utilizes a transformer as a secondary encoder after the CNNs encoder. The reason for this design is two-fold. Firstly, due to the lack of inductive bias in transformer models, encoding image feature with CNN networks leads to superior global feature modeling. Secondly, performing global self-attention on voxels in high-resolution medical images is computationally intensive. By using the CNN encoder first, the computational workload can be significantly reduced \cite{li2023transforming}. One early use of vanilla transformer blocks for medical segmentation is the TransUNet~\cite{chen2021transunet}, which used 12 2D transformer layers for encoding bottleneck features. TransUNet++~\cite{wang2022multiscale}, AFTer-UNet~\cite{yan2022after}, TransClaw~\cite{chang2021transclaw}, Ds-TransUNet~\cite{lin2021ds}, TransAttUNet~\cite{chen2021transattunet} and GT-UNet~\cite{li2021gt} improved the self-attention blocks and achieved promising performance in CT segmentation. In addition,  TransBTS~\cite{wang2021transbts}, CoTr~\cite{xie2021cotr}, and TrasnBridge~\cite{deng2021transbridge} explored variant modules such as deformable transformer blocks for 3D image segmentation tasks. Later, SegTrans~\cite{li2021medical}, MT-UNet~\cite{wang2021mixed} introduced squeeze and expansion mechanisms and mixed structure for modeling context affinities. BAT~\cite{wang2021boundary} and Poly-PVT~\cite{dong2021polyp} used grouping or boundary-aware designs to improve transformer robustness with cross-slice attention. 

The third design utilizes both transformer and CNNs encoders in parallel, which are also called fusion models. This design aims to take the global and local information from the transformer and CNNs encoder, respectively, for better representation learning. The encoded representations by two encoders are then fused into a single decoder. TransFuse~\cite{zhang2021transfuse}, and FusionNet~\cite{meng2021exploiting} are pioneering works that benefit from learning global and local features. The PMtrans~\cite{zhang2021pyramid} and X-Net~\cite{li2021x} introduce a multi-branch pyramid and a dual encoding network which demonstrate leading results on pathology images. MedT~\cite{valanarasu2021medical} and Ds-TransUNet~\cite{lin2021ds} proposed a CNN global branch and a local transformer branch with an axial self-attention module. With a fusion model, input medical images are split into both whole feature and non-overlapping patches followed by two encoder branches. With fusion designs, model complexities are commonly large due to the additional encoding branches, which is a disadvantage of these models. To the best of our knowledge, no fusion model has been proposed for volumetric medical image segmentation.

Recently, scientists have investigated the full adoption of transformer models for medical image segmentation. There are challenges in using pure transformer models, especially for 3D images, due to the limitation of inductive bias and the high complexity of transformers. Swin UNet~\cite{cao2021swin} is a pure transformer model designed for 2D medical images. It adopted the "U-shape" architecture and used a skip connection that connected the encoded features to the transformer decoder. D-Former~\cite{wu2022d} utilized dynamic position encoding blocks and local scope modules for improving local feature representation learning. MISSformer~\cite{huang2021missformer} is a pure transformer network with feed-forward enhanced blocks in its transformer modules. This design leveraged long-range dependencies with local features at different scales. The nnFormer~\cite{zhou2021nnformer} is another promising network that used 3D transformers and combined encoder and decoder with self-attention operations. nnFormer incorporated a skip attention mechanism to replace simple skip connections, which outperformed CNN-based methods significantly. Though the use of pure transformers as the model is more intuitive and better for design consistency; Yet, there are still uncharted areas using self-attention in the decoder. High model complexity can cause unsatisfied robustness and is challenging to explore in 3D context. 

Pre-training transformers with a large-scale dataset are of potential value to boost transformer model performance~\cite{dosovitskiy2020image}. Empirical studies~\cite{zhai2022scaling} show that the transformer model can have better scalability when more data are fed. In the medical domain, researchers have explored self-supervised pre-training approaches with CNNs~\cite{zhou2021models}. More recently, pre-training 3D transformers~\cite{tang2022self} for radiological images have been presented. Furthermore, uniformed pre-training frameworks~\citep{xie2021self,xie2021unified} are shown to construct teacher-student models for medical data. However, the use of pre-training is computationally exhaustive. In this paper, we aim to simplify and evaluate the effect of the pre-training framework with empirical studies. 

\noindent\textbf{Hierarchical Feature Aggregation.} The aggregation of multi-level features could improve segmentation results by merging the features extracted from different layers. Modeling hierarchical features, such as U-Net~\cite{cciccek20163d} and pyramid networks~\cite{roth2018multi}, multi-scale representations are leveraged. The extended feature pyramids compound the spatial and semantic information through two structures, iterative deep layer aggregation which fuses multi-scale information as well as deep hierarchical aggregation which fuses representations across channels. In addition to a single network, nested UNets~\cite{zhou2018unet++}, nnUNets~\cite{isensee2021nnu}, coarse-to-fine~\cite{zhu20183d} and Random Patch~\cite{tang2021high} suggest multi-stage pathways enrich the different semantic levels of features progressively with cascaded networks. Different from the above CNN-based methods, we explore the use of data-efficient transformers for modeling hierarchical 3D features by block aggregation.
\begin{figure*}[h!]
\centering
\includegraphics[width=\textwidth]{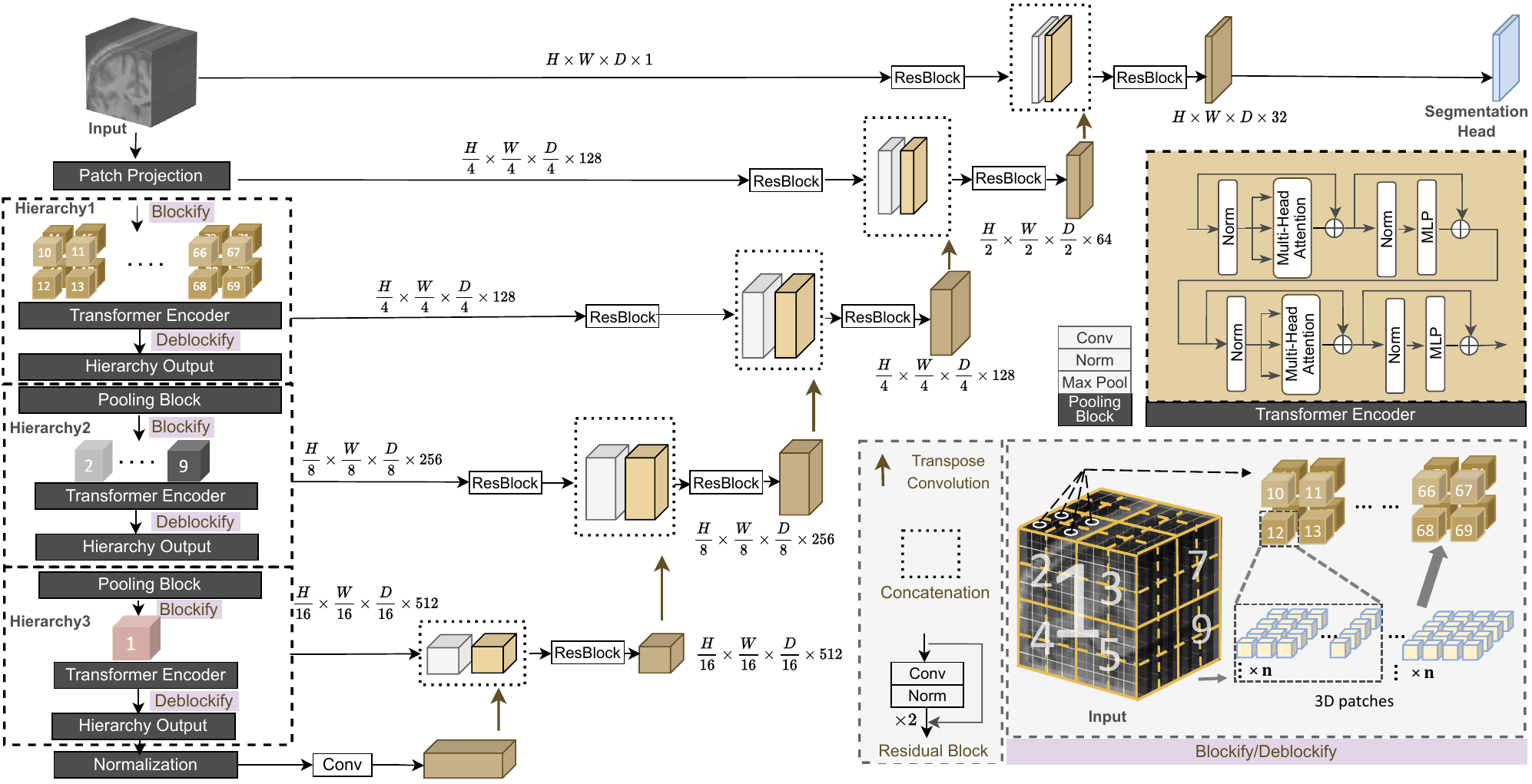}

\caption{Overview of the proposed UNesT with the hierarchical transformer encoder. Input image volumes are embedded into patches. In each hierarchy, patch embeddings are downsampled and blockified before being fed into the transformer encoder. Outputs are deblockified back to the volume plane. Each hierarchy output is connected with the decoder through skip connection. }

\label{fig:method}
\end{figure*}
\section{Method}
%
\subsection{Hierarchical Transformer Encoder}

The overall UNesT architecture is shown in Fig.~\ref{fig:method}. The input image is a sub-volume $\mathcal{X} \in \mathbb{R}^{H\times{W}\times{D}\times{C}}$ and the volumetric embedding token has a patch size of $S_h \times S_w \times S_d \times C$. 3D tokens are projected onto a size of $\frac{H}{S_h} \times \frac{W}{S_w} \times \frac{D}{S_d} \times C^{'}$ in the patch projection step, where $C^{'}$ is the embedded dimension. Following the motivation in \cite{zhang2022nested} for efficient non-local communication, all projected sequences of embeddings are partitioned to blocks (blockify) with a resolution of $\mathcal{X} \in \mathbb{R}^{b \times T \times{n} \times C^{'}}$, where $T$ is the number of blocks at the current hierarchy, $b$ is the batch size, $n$ is the total length of sequences, as shown in Fig.~{\ref{fig:method}}. The blocks are non-overlapping and $n$ remains the same in different hierarchies.
The dimensions of the embeddings follow $T \times{n} = \frac{H}{S_h} \times \frac{W}{S_w} \times \frac{D}{S_d}$. Each block is fed into sequential of transformer layers\cite{sharir2021image} separately, which consist of the canonical multi-head self-attention (MSA), multi-layer perceptron (MLP) with skip connection~\cite{he2016identity}, and layer normalization (LN)~\cite{ba2016layer}. We add learnable position embeddings to sequences for capturing spatial relations before the blocked transformers. The output of transformer encoder is computed as follows:

\begin{equation}
\begin{array}{l}
\hat{{z}}^{t}=\text{MSA\textsubscript{HRCHY\textsubscript{l}}}(\text{LN}({z}^{t-1}))+{z}^{t-1}, t=1...L \\
{z}^{t}=\text{MLP}(\text{LN}(\hat{{z}}^{t}))+\hat{{z}}^{t}, t=1...L \\

\end{array}
\label{eq:eq1}
\end{equation}
where MSA\textsubscript{HRCHY\textsubscript{l}} denotes the multi-head self-attention layer of hierarchy $l$, $\hat{z}^{t}$ and $z^{t}$ are the output representations of MSA and MLP and L denotes the number of transformer layers. $z^{t-1}$ denotes the input of the transformer encoder, as shown in Fig~{\ref{fig:method}}, after undergoing layer normalization and MSA, $z^{t-1}$ is added to the output via a skip connection to produce {$\hat{{z}}^{t}$}. The result is then passed through another layer hl{normalization} and a MLP. Finally, {$\hat{{z}}^{t}$} is added to the final result to generate the output of a transformer layer, denoted by {${z}^{t}$}, which serves as the input to the subsequent transformer layer.

In practice, MSA\textsubscript{HRCHY\textsubscript{l}} is applied in parallel to all partitioned blocks:
\begin{equation}
\small\begin{array}{l}
\text{MSA\textsubscript{HRCHY\textsubscript{l}}(Q, K, V)} =  \text{Stack}(\text{BLK}\textsubscript{1},..., \text{BLK}\textsubscript{T})\\ \text{BLK}=\text{{{MSA}}}(Q, K, V)= \text{Stack}(\text{Softmax}(\frac{Q_iK_i^{\top}}{\sqrt{\sigma}})V_i)W^o, i=1...H,
\end{array}
\label{eq:eq2}
\end{equation}
where $Q, K, V$ denotes queries, keys, and value vectors in the multi-head attention, and H represents the total heads in the MSA. In each block, $Q, K, V$ have the dimension of $\sigma$. Dot-product attention is applied between $Q$ and $K^T$ to get an attention matrix of size $\sigma \times \sigma$. To overcome the issue that when $\sigma$ has a large value, the dot product between $Q$ and $K$ becomes magnified causing the softmax function to produce extreme value, the dot products is scaled by $\frac{1}{\sqrt{\sigma}}$~\mbox{\cite{vaswani2017attention}}. $V$ is then multiplied with the attention matrix to get the final output with size $\sigma$. The computation of each head is performed in parallel and then combined through concatenation. The final result is then reshaped with matrix $W \in \mathbb{R}^{H \cdot \sigma \times d_{out}}$ to match the output dimension. As previously stated, each block shares a common size of $b \times{n} \times C^{'}$ within the same hierarchy so that the MSA output of each block has the same size. The outputs of each block are concatenated to obtain the final results, which represent the MSA of that particular hierarchy.  All blocks at each level of the hierarchy share the same parameters given the input $\mathcal{X}$, which leads to hierarchical representations without increasing complexity. 

\subsection{3D Block Aggregation}
 We extend the spatial nesting operations in \cite{zhang2022nested} to 3D blocks to form a local attention hierarchical design. Different to~\citep{liu2021swin,tang2022self}, which utilizes global attention among "shift windows". In our design, transformer encoders are applied to each volume block separately to achieve local attention, with each block being modeled independently. Information across blocks is communicated by the aggregation module. This design leads to reduced computational complexity and improved data efficiency.
 
In the first hierarchy, suppose the input feature size is $H^{'} \times W^{'} \times D^{'} \times C^{'}$. The input feature is blockified into $T$ blocks with the aforementioned size of $T \times{n} \times C^{'}$. After the transformer encoder, the blocks are transformed back to the feature map with size $H^{'} \times W^{'} \times D^{'} \times C^{'}$, which serves as the input of the next hierarchy. In the following hierarchy,  the input feature downsamples by a factor of 2 for each dimension and transformed to embedded dimension before blockify with a pooling block consisting of a convolutional layer, a normalization layer and a max pooling layer to build multi-scale feature maps as in~\mbox{\citep{ronneberger2015u,liu2021swin}} for better representation learning. Applying the pooling block before blockify facilitates the information exchange between blocks and enables non-local communication because it allows convolution and pooling to be performed on the spatial area that belongs to different blocks after blockifying. The pooling block reduces the feature size by $2 \times 2 \times 2$, and the sequence length  $n$ remains the same, the number of blocks T reduces by a factor of 8 in each hierarchy as a result. Consequently, in the subsequent level of the hierarchy, the feature maps are blockified to the size of $\frac{T}{8} \times{n} \times C^{''}$, where $C^{''}$ represents the embedded dimension in that hierarchy. The unblocked feature maps after the transformer encoder have a size of $\frac{H^{'}}{2} \times \frac{W^{'}}{2} \times \frac{D^{'}}{2} \times C^{''}$. This process continues until the number of blocks $T$ reach 1.
 
 In our model design, there are three hierarchies which result in a total number of 64, 8, and 1 block in each hierarchy. In the volumetric plane, the encoded blocks are merged among adjacent block representations, as it shown in Fig.~{\ref{fig:method}}. The design and use of the aggregation modules in the 3D scenario leverage local attention and improved data efficiency which we demonstrate in our ablation studies.
 
\subsection{Decoder}
To better capture localized information and further reduce the effects of lacking inductive bias in transformers, we use a hybrid design with a convolution-based decoder for segmentation.

We use a patch size of $4 \times 4 \times 4$ in our encoder. The patch embedded dimension is set to 128 so that the feature size is $\frac{H}{4} \times \frac{W}{4} \times \frac{D}{4} \times 128$ after patch projection. The 3 hierarchies have a number of transformer layers (depth) of 2, 2, and 8 and embedded dimensions (width) of 128, 256, and 512, respectively. The feature size at the end of each hierarchy is$\frac{H}{4\times2^i} \times \frac{W}{4\times2^i} \times \frac{D}{4\times2^i} \times C$ where $i=0,1,2$ and $C=128,256,512$, as shown in Fig.~\ref{fig:method}. The feature map from the last hierarchy is fed into a layer normalization layer to generate the transformer encoder output. 

As previously mentioned, the feature size is reduced by a factor of 2 in each dimension at each level of the hierarchy, resulting in the generation of multi-resolution feature maps. Inspired by the U-shape models \cite{ronneberger2015u} which utilize multi-scale strategy, we merge the multi-resolution features which is the output of each hierarchy with the decoder with skip connections followed by convolutional layers.

The bottleneck is generated by feeding the output of the last hierarchy to a layer normalization followed by a $3 \times 3 \times 3$ convolutional layer. We upsample the bottleneck by applying a transpose convolutional layer. The output of the transposed convolution is concatenated with the prior hierarchical representations and fed into a residual block consisting of two $3 \times 3 \times 3$ convolutional layers, each followed by an instance normalization \cite{ulyanov2016instance} layers (Fig.~{\ref{fig:method}}). In each hierarchy, excluding the last one, we have incorporated a residual block with the aforementioned layers in the skip connection between the hierarchy's output (encoder) and decoder. Since we believe these residual blocks can help minimize the semantic gap between the features from the transformer encoder and the CNN decoder. However, since the bottleneck in our architecture involves transforming the output of the last layer using a normalization layer and a convolutional layer, the semantic meaning is expected to be similar to that of the last hierarchy's output. Therefore, we choose not to include a residual block in the skip connection of the last hierarchy.
The processed feature maps from the encoder are then concatenated with the feature maps from lower hierarchies or bottleneck upsampled by transposing convolutional layers. This merged feature map is then passed through another residual block with the layers mentioned earlier to merge the information from both the encoder and decoder. To enhance the semantic information, we apply the aforementioned residual block to both the input image and the features obtained after the path projection. The resulting features were then passed to the decoder through the skip connection, as illustrated in Fig.~{\ref{fig:method}}.
 The segmentation mask is acquired by $1 \times 1 \times 1$ convolutional layer with a softmax activation function. Compared to some prior related works such as TransBTS~\cite{wang2021transbts} and CoTr~\cite{xie2021cotr}, our design employs the hierarchical transformer directly on images and extracts representations at multiple scales without convolutional layers. 

\section{Experiments}
\subsection{Dataset}
\label{dataset}
\noindent\textbf{Whole Brain Segmentation Dataset.}
Training and testing data are MRI T1-weighted (T1w) 3D volumes from 10 different sites. The training set consists of 50 scans from the Open Access Series on Imaging Studies (OASIS) \cite{marcus2007open} dataset which is manually traced to 133 labels based on the BrainCOLOR protocol \cite{klein2010open} by Neuromorphometrics Inc. The size of the data is 256 $\times$ 256 $\times$ [270,334] with 1 mm isotropic spacing. The testing cohort contains Colin27 (Colin) T1w scan~\cite{aubert2006new} and 13 T1w MRI scans from the Child and Adolescent Neuro Development Initiative (CANDI)~\cite{kennedy2012candishare} dataset. The Colin dataset contains one high-resolution scan averaging from 27 scans of the same subject. The label is manually traced to 130 labels based on BrainCOLOR protocol. The size of the scan is 362 $\times$ 362 $\times$ 434  with 0.5mm isotropic spacing. The CANDI dataset is manually traced to 130 labels following the BrainCOLOR protocol. The size of the scan is 256 $\times$ 256 $\times$ 128 with spacing of 0.94mm $\times$ 0.94mm $\times$ 1.5mm. A detailed class name and the 3 classes not labeled in the test sets can be found in Table~\ref{tab:brainlabel} in the supplementary material. The CANDI dataset contains a different age group (5-15 years old) compared to the OASIS training cohort (18-96 years old), which allowed assessment of different populations. Following the same practice in \cite{huo20193d}, we use auxiliary labels comprising of 4859 T1w MRI scans from eight different sites whose labels are generated by using an existing multi-atlas segmentation pipeline~\cite{asman2014hierarchical} to pre-train the model and finetune the pre-trained model with the 50 manually traced data from the OASIS dataset. A detailed summary of the 4859 multi-site images is shown in Table~\ref{tab:dataset}. 


\begin{table*}[ht]
\centering
\caption{Data summary of the 4859 multi-site images.}
\label{tab:dataset}
\begin{tabular}{lll}
\hline
Study Name                                         & Website                                     & Images \\ \hline
Baltimore Longitudinal Study of Aging (BLSA)       & www.blsa.nih.gov                            & 614    \\
Cutting Pediatrics                                 & vkc.mc.vanderbilt.edu/ebrl                  & 586    \\
Autism Brain Imaging Data Exchange (ABIDE)         & fcon\_1000.projects.nitrc.org/indi/abide    & 563    \\
Information Extraction from Images (IXI)           & www.nitrc.org/projects/ixi\_dataset         & 541    \\
Attention Deficit Hyperactivity Disorder (ADHD200) & fcon\_1000.projects.nitrc.org/indi/adhd200  & 950    \\
Open Access Series on Imaging Study (OASIS)        & www.oasis-brains.org                        & 312    \\
1000 Functional Connectome (fcon\_1000)            & fcon\_1000.projects.nitrc.org               & 1102   \\
Nathan Kline Institute Rockland (NKI\_rockland)    & fcon\_1000.projects.nitrc.org/indi/enhanced & 141    \\ \hline
\end{tabular}
\end{table*}

\begin{figure}
\centering
\includegraphics[width=0.45\textwidth]{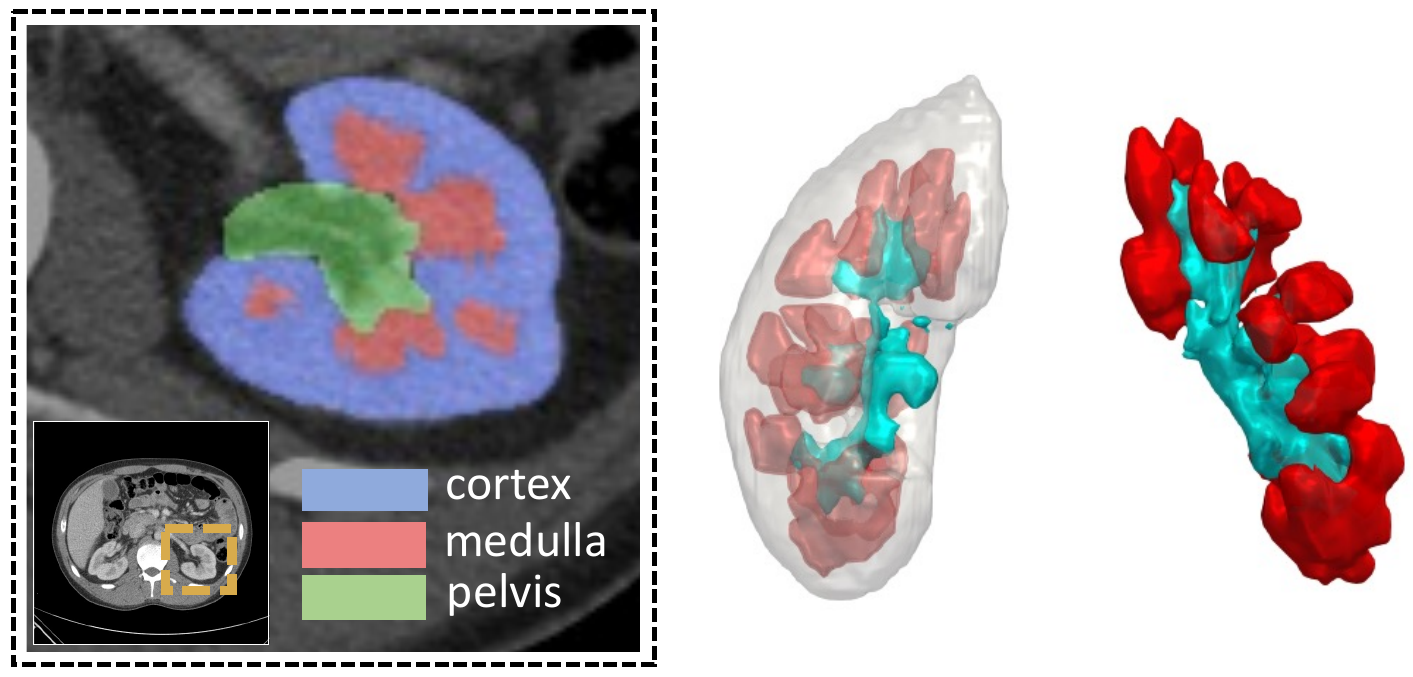}
\caption{Visual and 3D illustration of the kidney components.}
\label{fig:renal}
\end{figure}

\noindent\textbf{Renal Substructure Dataset.}
We construct an internal cohort of the renal substructures segmentation dataset with 116 subjects imaged under institutional review board (IRB) approval (IRB \#131461). Cortex, medulla, and pelvicalyceal systems are labeled in the dataset (Fig.~\ref{fig:renal}). Data with ICD codes related to kidney dysfunction are excluded since they could potentially influence kidney anatomy. The left and right renal structures are outlined manually by three interpreters under the supervision of clinical experts. The renal columns are included in the cortex label. The medulla is surrounded by the cortex, and the pelvicalyceal systems contain calyces and pelvis that drain into the ureter. All manual labels are verified and corrected independently by expert observers. For the test set of 20 subjects, we perform a second round of manual segmentation (interpreter 2) to assess the intra-rater variability and reproducibility. The image size of each scan is 512 $\times$ 512 $\times$ [90,131] with spacing of [0.54,0.98]mm $\times$ [0.54,0.98]mm $\times$ 3.0mm.

\noindent\textbf{Multi-organ Segmentation (BTCV) Dataset.} We evaluate model generalizability with the Beyond The Cranial Vault (BTCV) dataset. It is comprised of 100 de-identified contrast-enhanced CT volumes with 13 labeled anatomies, including spleen, right kidney, left kidney, gallbladder, esophagus, liver, stomach, inferior vena cava (IVC), portal and splenic veins (PSV), pancreas, right and left adrenal gland. The image size of each scan is 512 $\times$ 512 $\times$ [80, 255] with the spacing of [0.54, 0.98]mm $\times$ [0.54, 0.98]mm $\times$ [2.5, 7.0]mm. 50 scans are publicly available in the MICCAI 2015 Multi-atlas Labeling Challenge \cite{landman2015miccai}, in which 20 scans are used for public testing.

\noindent\textbf{KiTS19.}
To further validate the generalizability of the proposed method for characterizing renal tissues, we apply the model to the public KiTS19 dataset. The KiTS19~\cite{heller2021state} task focuses on the whole kidney and kidney tumor segmentation. Images and labels from 210 subjects are publicly available. The image size of each scan is 512 $\times$ [512, 796] $\times$ [29, 1059] with spacing of [0.44, 1.04]mm $\times$ [0.44, 1.04]mm $\times$ [0.5, 5.0]mm.

\noindent\textbf{Brats.}
The BraTS 2021 dataset contains 1251 subjects, and each scan is associated with 4 MRIs: 1) native (T1) and 2) post-contrast T1-weighted (T1Gd), 3) T2-weighted (T2), and 4) T2 Fluid-attenuated Inversion Recovery (T2-FLAIR). Each subject's images are registered and resampled to $1.0 \times 1.0 \times 1.0$ mm isotropic resolution. The input 3D volumes are of size  $240 \times 240 \times 155$. 

\subsection{Implementation Details}
\label{details}
To eliminate the impact of data difference on the final performance, all baseline models undergo the same data augmentation and pre-processing steps, except of the nnUNet framework-based methods. Experiments are implemented in Pytorch and MONAI\footnote{\href{https://monai.io}{https://monai.io/}}. All segmentation models are trained with a single Nvidia RTX 5000 16G GPU with an input volume size of $96 \times 96 \times 96$. We follow the initial learning rate and weight decay configurations used in each respective baseline, provided that the original baseline was tested on the same task. If the baseline does not test on that task, for nn-UNet based method (nn-UNet and CoTr), we keep the initial settings since the training planner in the code has the original learning rate set as default. For the other models, we adopt an identical learning rate of 1e-4 and weight decay of 1e-4. The rationale behind this decision is twofold. Firstly, these baselines' initial learning rate and weight decay settings match or closely resemble this setting. Secondly, we utilize a cosine scheduler with 500 step warm-up for all the models which will adjust the learning rate based on the models accordingly.

\subsubsection{Whole Brain Segmentation}
\begin{figure}
\centering
\includegraphics[width=0.45\textwidth]{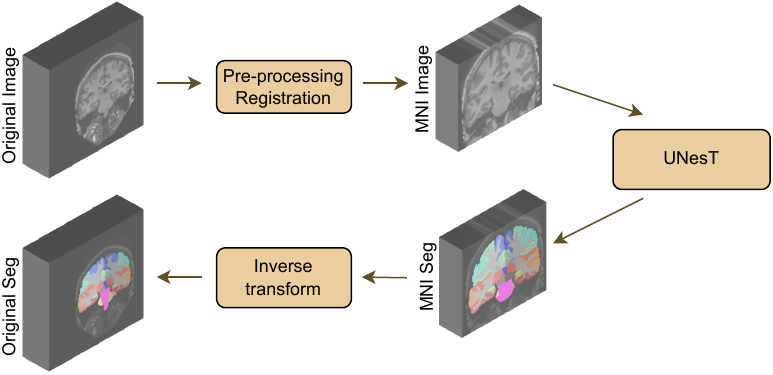}
\caption{Overview of the workflow for the whole brain segmentation task. Original images are pre-processed and registered to the MNI space before feeding into the networks. Model outputs that are in MNI space are transformed back to the original space to get the final predictions.}
\label{fig:workflow}
\end{figure}

During pre-training with auxiliary labels, the learning rate is initialized to 0.0001 with weight decay of $1e^{-4}$ to train for 200K iterations. During finetuning, the learning rate is set to $1e^{-5}$ to train for 50K iterations. As shown in Fig.~\ref{fig:workflow}, all data are registered to the MNI space using the MNI305~\cite{evans19933d} template and preprocessed following the method in \cite{huo20193d}. All processed images have a size of 172 $\times$ 220 $\times$ 156 with isotropic spacing of 1mm. Registered input images are randomly cropped to the size of 96 $\times$ 96 $\times$ 96 during the online augmentation. We use a five-fold cross-validation strategy during finetuning. The best-performing model in each fold is selected to test on the external testing set and ensembled to get the final prediction in the MNI space. Predictions in MNI space are inverse transformed to the original space using NiftyReg \cite{ourselin2001reconstructing} for evaluation (Fig.~\ref{fig:workflow}). No data augmentation is used in all the experiments due to the negative impact on model performance observed during our experiments.  Segmentation performances are evaluated using Dice similarity coefficient (DSC) and symmetric Hausdorff Distance (HD).
\subsubsection{Renal Substructures Segmentation}
 Five-fold cross-validation is used for all experiments on 96 subjects, while 20 subjects are used for held-out testing. The five-fold models' ensemble is used for inference and evaluating test set performance. For experiment training, we used 1) a CT window range of [-175, 275] HU; 2) scaled intensities of [0.0,1.0] with 1.0mm isotropic spacing. The learning rate is initialized to 0.0001, followed by a weight decay of $1e^{-4}$ for 50K iterations. Common data augmentation such as random flip, rotation, and change of intensity are applied with the probability of 0.1. For fair comparison and direct evaluation of the effectiveness of models, no pre-training is performed for all segmentation tasks. Segmentation results are evaluated with DSC and HD. We conduct volumetric analyses on kidney components in terms of R squared error, Pearson R, absolute deviation of volume, and the percentage difference between the proposed method and manual label. 

\subsubsection{Multi-organ Segmentation}
80 subjects are used for training/validation and 20 are used for testing. The images are resampled to 1.5mm $\times$ 1.5mm $\times$ 2.0mm. We perform the same data augmentation as the renal substructure segmentation. The learning rate is initialized to 0.0001 followed by a weight decay of $1e^{-4}$ for 100K iterations. Segmentation results are evaluated with DSC. 

\subsubsection{Kidney and Kidney Tumor Segmentation}
We perform five-fold cross-validation experiments on 210 subjects and show DSC results of the held-out 20\%. The experiments have the same settings as the renal substructures dataset.

\subsubsection{Brain Tumor Segmentation}
Following the same data split of Swin UNETR \mbox{\cite{tang2022self}}, SegResNet \cite{myronenko20193d}, and nnUNET \mbox{\cite{isensee2021nnu}}, we train our method with five-fold cross-validation with a ratio of 0.8 and 0.2.

\begin{table*}[ht]
\centering
\caption{Performance comparison for the whole brain segmentation task. Overall UNesT achieved state-of-the-art performance on the whole brain segmentation task. The number of parameters and GFLOPs (with a single input volume of $96 \times 96 \times 96$ for the transformer-based models) are shown. "$\times 27$" in SLANT27 represents that 27 of the same models are used. inf denotes part of data in the testing datasets have infinite HD. Notes: the FLOPs for SLANT27 are calculated based on input size of 96 $\times$ 128 $\times$ 88 as designed in the paper \cite{huo20193d}.}
\label{tab:brainresults}
\begin{adjustbox}{width=0.9\textwidth}
\begin{tabular}{l|cc|cccc}
\hline
\multirow{2}{*}{Method}                                  & \multirow{2}{*}{\#Param} & \multicolumn{1}{l|}{\multirow{2}{*}{FLOPs(G)}} & \multicolumn{2}{c}{Colin}                                           & \multicolumn{2}{c}{CANDI}                                           \\ \cline{4-7} 
                                                         &                          & \multicolumn{1}{l|}{}                          & DSC                              & HD                               & DSC                              & HD                               \\ \hline
nnUNet~\cite{isensee2021nnu}      & 30.7M            & 358.6                                          & 0.7168                           & 10.7321                          & 0.4337                           & inf                              \\
TransBTS~\cite{wang2021transbts}  & 33.0M                    & 111.9                                          & 0.6537                           & inf                              & 0.6043                           & inf                              \\
nnFormer~\cite{zhou2021nnformer}  & 158.9M                   & 920.1                                          & 0.7113                           & 10.2755                          & 0.6393                           & inf                              \\
CoTr~\cite{xie2021cotr} &42.0M &328.0 &0.7209 &10.3194 &0.6908 &inf\\
UNETR~\cite{hatamizadeh2022unetr} & 92.6M                    & 268.0                                           & 0.7320                           & 10.3834                          & 0.6851                           & 11.1972                          \\
SwinUNETR~\cite{tang2022self}     & 62.2M                    & 334.9                                          & 0.6854                           & 22.0389                          & 0.6537                           & 34.3980                          \\
SLANT27~\cite{huo20193d}          & 19.9M $\times$ 27        & 2051.0 $\times$ 27                                              & 0.7264                           & \textbf{9.9061} & 0.6968                           & 8.8851                           \\ \hline
UNesT                                                    & 87.3M                    & 261.7                                          & \textbf{0.7444} & 11.0081                          & \textbf{0.7025} & \textbf{8.8417} \\ \hline
\end{tabular}
\end{adjustbox}
\end{table*}

\begin{table}[]
\caption{Single model performance for the whole brain segmentation task. Models are trained with the same training/validation data.}
\label{tab:whole_single}
\resizebox{\linewidth}{!}{%
\begin{tabular}{l|cccc}

\hline
                               & \multicolumn{2}{c}{Colin} & \multicolumn{2}{c}{CANDI} \\ \cline{2-5} 
\multirow{-2}{*}{Method}        & DSC                     & HD                      & DSC                     & HD                      \\ \hline
nnUNet~\cite{isensee2021nnu}    & 0.7062                  & 14.0101                 & 0.3930                  & inf                     \\
TransBTS~\cite{wang2021transbts}  & 0.6542                  & inf                     & 0.5991                  & inf                     \\
nnFormer~\cite{zhou2021nnformer}  & 0.7007                  & 10.423                  & 0.6420                  & inf                     \\
CoTr~\cite{xie2021cotr}          & 0.7268                  & 10.2561                 & 0.6923                  & inf                     \\
UNETR~\cite{hatamizadeh2022unetr} & 0.7328                  & 10.216                  & 0.6810                  & 13.3172                 \\
SwinUNETR~\cite{tang2022self}    & 0.6853                  & 21.4812                 & 0.6536                  & 34.5212                 \\
SLANT~\cite{huo20193d}                                                  & 0.7301                  & \textbf{9.9470}         & 0.6977                  & 9.5000                  \\\hline
UNesT                                                   & \textbf{0.7467}         & 11.0358                 & \textbf{0.7022}         & \textbf{8.8902}         \\ \hline
\end{tabular}%
}
\end{table}

\section{Results}
We evaluate the UNesT performance against recent convolutional-~\cite{isensee2021nnu} and transformer-based~\citep{wang2021transbts,zhou2021nnformer,hatamizadeh2022unetr,tang2022self} 3D medical segmentation baselines. UNesT presents distinguished results on the task of whole brain segmentation with 133 tissue classes. Next, we perform experiments on the first kidney substructures CT dataset. We further validate model generalizability with the publicly available BTCV, KiTS19, and BraTs2021 datasets.

\subsection{Whole Brain Segmentation}
\begin{figure}
\centering
\includegraphics[width=0.4\textwidth]{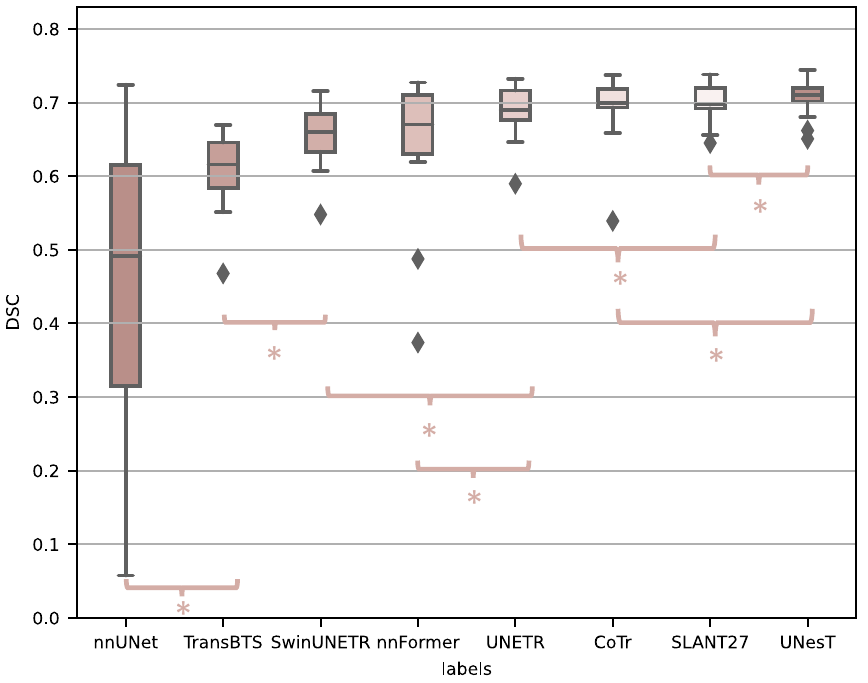}
\caption{Quantitative results of the whole brain segmentation on the testing data. SLANT27 shows the smallest variation among the other baselines. UNesT achieves the overall best performance. Compared with SLANT27, UNesT further reduces the variation with improved median and quartiles of the DSC. * indicates statistically significant ($p < 0.05$) by Wilcoxon signed-rank test. Detailed quantitative performance comparison of 130 classes is shown in Fig~{\ref{fig:133boxplot}} in the supplementary material.}
\label{fig:brain_boxplot}
\end{figure}

\begin{figure*}[h!]
\centering
\includegraphics[width=\textwidth]{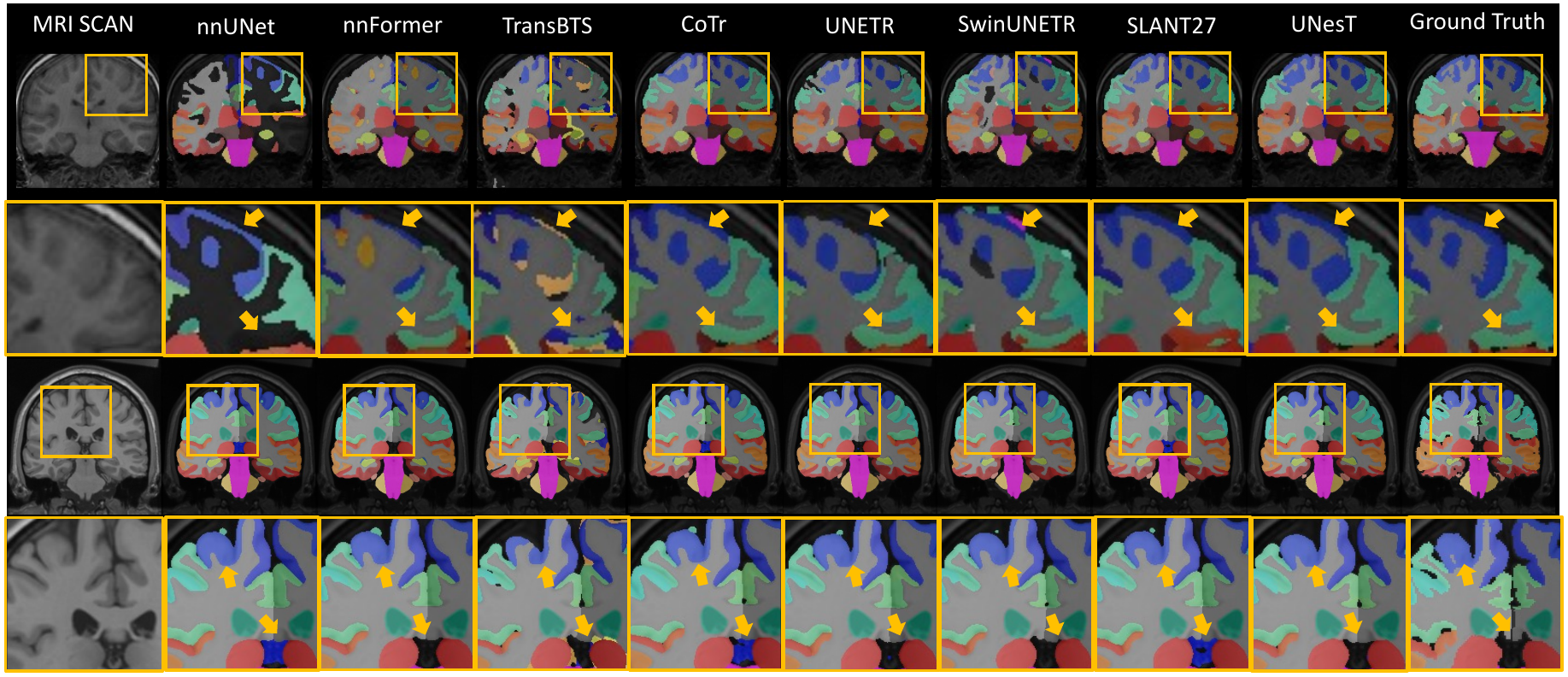}
\caption{Qualitative results of whole brain segmentation on the CANDI dataset (top 2 rows) and Colin dataset (bottom 2 rows). Boxed areas are enlarged in the lower row. Differences are emphasized with the orange arrow. UNesT shows better captures the boundary and correctly segments the tissues.}
\label{fig:brain_vis}
\end{figure*}

A detailed comparison of quantitative performance is shown in Table~\ref{tab:brainresults} and Fig.~\ref{fig:brain_boxplot}. The qualitative performance is shown in Fig.~\ref{fig:brain_vis}. All the models are pre-trained with 4859 auxiliary pseudo labels and are finetuned with 50 manually traced labels from OASIS in the 5-fold ensemble setting. We first compare the proposed UNesT model with nnUNet~\cite{isensee2021nnu} and several transformer-based methods. Most of the methods have infinite HD on the CANDI dataset associated with 0.43 to 0.69 DSC score indicating those methods fail to predict all of the 130 classes in the external testing set. UNETR performs the best among these widely used 3D medical image segmentation methods. Compared with UNETR, UNesT improves the performance in the Colin (from 0.7320 to 0.7444) and the CANDI (from 0.6851 to 0.7025) dataset by a margin. SLANT27~\cite{huo20193d}, the prior state-of-the-art method, divides the whole brain into 27 parts and ensembles 27 tiled 3D-UNet \cite{cciccek20163d} for the final predictions. Within the same 5-fold ensemble settings, UNesT ensembled with 5 models outperforms SLANT27 ensembled with 135 models in terms of DSC in both Colin (0.7444 vs. 0.7264) and CANDI (0.7025 vs. 0.6968) dataset and achieves the state-of-the-art performance. UNesT achieves significant improvement on the test set compared to SLANT27 with $p < 0.05$ under Wilcoxon signed-rank test and further reduces the variation of DSC score distribution with tighter quartiles (Fig.~\ref{fig:brain_boxplot}). In Fig.~\ref{fig:brain_vis}, we show UNesT has better captures on the boundary and correctly segments brain tissues. As the external testing set represents a high resolution and different age population cohort, we show that our method can generalize learned knowledge to different populations.

\subsection{Characterization of Renal substructures}

\begin{figure*}[h!]
\centering
\includegraphics[width=0.98\textwidth]{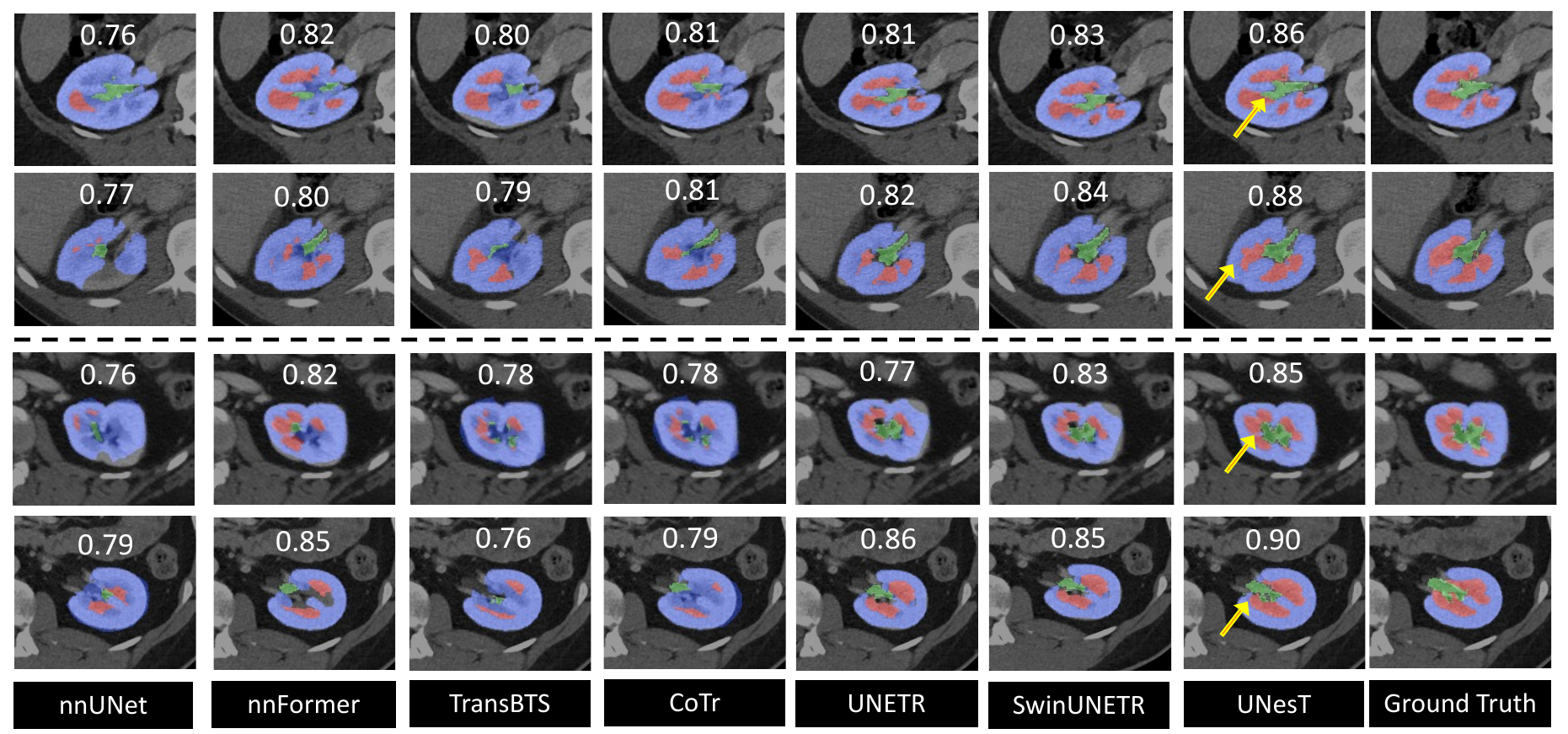}
\caption{Qualitative comparisons of representative renal sub-structures segmentation on two right (top) and two left (bottom) kidneys. The average DSC is marked on each image. UNesT shows distinct improvement on the medulla (red) and pelvicalyceal system (green) against baselines. Comparisons with different baselines including the ViT and CNN hybrid approaches.}
\label{fig:fig3}
\end{figure*}

\begin{figure*}[h!]
\centering
\includegraphics[width=\textwidth]{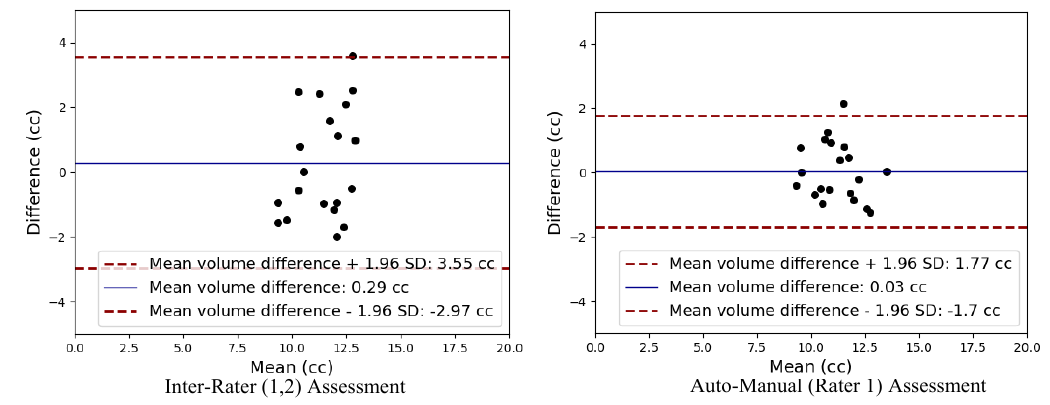}
\caption{The Bland-Atman plots compare the medulla volume agreement of inter-rater and auto-manual assessment. We show cross-validation on interpreter 1 and interpreter 2 manual segmentation on the same test set.  Interpreters present independent observation without communication. The auto-manual assessment shows the agreement between UNesT and interpreter 1 annotation. }
\label{fig:figA1}
\end{figure*}

\noindent {\bf Segmentation Results.}
Compared to canonical kidney studies using shape models or random forests in Table~\ref{tab:tab1}, deep learning-based methods improve the performance by a large margin from 0.7233 to 0.7991. Among the nnUNet~\cite{isensee2021nnu} and extensive transformer models, we obtain the state-of-the-art average DSC score of 0.8564 compared to the second-best performance of 0.8411 from SwinUNETR, with a significant improvement $p < 0.05$ under Wilcoxon signed-rank test. We observe higher improvement in smaller anatomies such as the medulla and collecting systems. We compare qualitative results in Fig.~\ref{fig:fig3}. Our method demonstrates the distinct improvement of detailed structures for the medulla and pelvicalyceal systems. Fig.~\ref{fig:figA1} shows that the proposed automatic segmentation method achieves better agreement compared to inter-rater assessment, 0.03 against 0.29 of mean difference indicating reliable reproducibility.

\noindent {\bf Volumetric Analysis.}
Table~\ref{tab:tab2} lists the volume measurement with the proposed method. The UNesT achieves an R squared error of 0.9348 on the cortex. The correlation performance metric with Pearson R achieves 0.9896 for the UNesT against the manual label on the cortex. Our method obtains 2.5259 with an absolute deviation of volumes. The percent difference in the cortex is 3.8411. We observe the same trend for the Medulla and Pelvicalyceal systems. Quantitative results show that our workflow can serve as the state-of-the-art volumetric measurement compared to the prior kidney characterization state-of-the-art~\cite{tang2021renal}. 

\begin{table*}[ht]
\centering
\scriptsize
\setlength{\tabcolsep}{2mm}
\renewcommand\arraystretch{1}
\caption{Segmentation results of the renal substructure on testing cases. The UNesT achieves state-of-the-art performance compared to prior kidney component studies and 3D medical segmentation baselines. * indicates statistically significant ($p < 0.05$) compared with the underlined performance by Wilcoxon signed-rank test.}
\begin{adjustbox}{width=0.95\textwidth}
\begin{tabular}{l|cccccc|cc}
\hline
\multirow{2}{*}{Method}   &\multicolumn{2}{c}{Cortex}  &\multicolumn{2}{c}{Medulla} &\multicolumn{2}{c}{Pelvicalyceal System} &\multicolumn{2}{|c}{Avg.} \\
\cline{2-9}
 &DSC &HD &DSC &HD &DSC &HD &DSC &HD\\
\hline
Chen et al.~\cite{chen2012automatic}   &0.7512 & 40.1947    &N/A &N/A    &N/A &N/A     &N/A &N/A    \\
Xiang et al.~\cite{xiang2017cortexpert}  &0.8196 & 27.1455    &N/A &N/A    &N/A &N/A     &N/A &N/A    \\
Jin et al.~\cite{jin20163d}  &0.8041 & 34.5170    &0.7186 &32.1059    &0.6473 &39.9125     &0.7233 &35.5118    \\
Tang et al.~\cite{tang2021renal} &0.8601 & 19.7508    &0.7884 &18.6030    &0.7490 &34.1723     &0.7991 &24.1754    \\
\hline
nnUNet~\cite{isensee2021nnu} &0.8915 & 17.3764    &0.8002 &18.3132    &0.7309 &31.3501     &0.8075 &22.3466    \\
TransBTS~\cite{wang2021transbts}	&0.8901 & 17.0213    &0.8013 &17.3084    &0.7305 &30.8745     &0.8073 &21.7347    \\
CoTr~\cite{xie2021cotr}	&0.8958 & 16.4904    &0.8019 &16.5934    &0.7393 &30.1282     &0.8123 &21.0707    \\
nnFormer~\cite{zhou2021nnformer}  &0.9094 & 15.5839    &0.8104 &15.9412    &0.7418 &29.4407     &0.8205 &20.3219    \\
UNETR~\cite{hatamizadeh2022unetr}  &0.9072 & 15.9829    &0.8221 &14.9555    &0.7632 &27.4703     &0.8308 &19.4696    \\
SwinUNETR~\cite{tang2022self}   &0.9182 &\textbf{14.0585} &0.8344 &11.9582 &0.7707 &14.6027 &\underline{0.8411} &13.5398 \\
\hline
UNesT 	&\textbf{0.9262} &14.4628    &\textbf{0.8471} &\textbf{8.3677}    &\textbf{0.7958} &\textbf{9.735}     &\textbf{0.8564*} &\textbf{10.1885}    \\
\hline
\end{tabular}
\end{adjustbox}
\label{tab:tab1}
\end{table*}
\begin{table*}[ht]
\centering
\scriptsize
\setlength{\tabcolsep}{2mm}
\renewcommand\arraystretch{1}
\caption{Comparison of volumetric analysis metrics between the proposed method and the state-of-the-art clinical study on kidney components.}
\resizebox{0.95\textwidth}{!}{
\begin{tabular}{l|cc|cc|cc}
\hline
\multirow{2}{*}{Metrics}   &\multicolumn{2}{c}{Cortex}  &\multicolumn{2}{c}{Medulla} &\multicolumn{2}{c}{Pelvicalyceal System} \\
\cline{2-7}
 &~\cite{tang2021renal}  &UNesT &~\cite{tang2021renal} &UNesT &~\cite{tang2021renal} &UNesT 
\\
\hline
R Squared		&0.9200 & 0.9348    &0.6652 &0.6850    &0.4586 &0.6126       \\
Pearson R			&0.9838 & 0.9896    &0.8156 &0.8428    &0.6772 &0.7454     \\
Absolute Deviation of Volume &3.0233 & 2.5259    &3.5496 &3.0293    &0.9443 &0.7410      \\
Percentage Difference			&4.8280 & 3.8411    &7.4750 &6.894    &19.0716 &12.0171   \\

\hline
\end{tabular}
}

\label{tab:tab2}
\end{table*}

\begin{figure*}[h!]
\centering
\includegraphics[width=\textwidth]{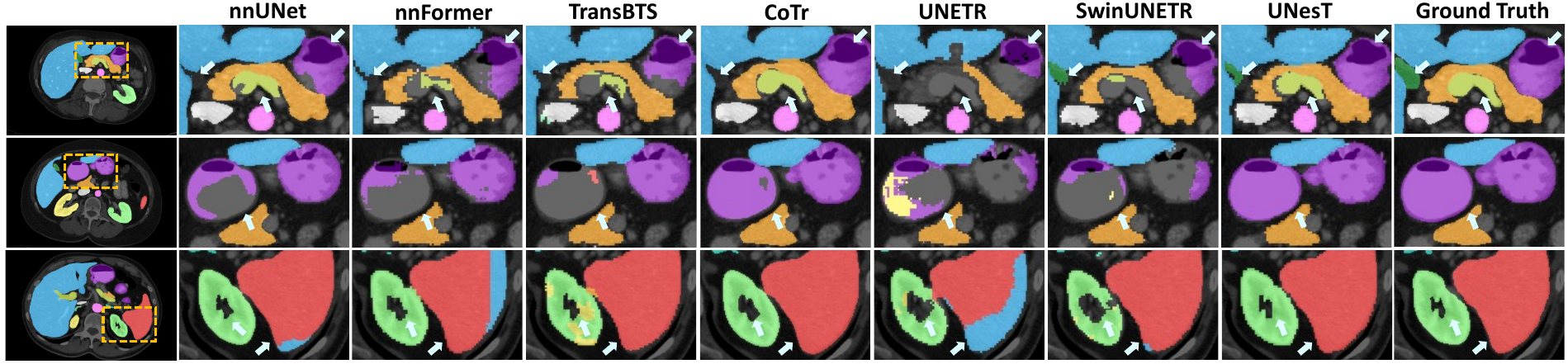}
\caption{Qualitative comparison between UNesT and baseline methods on the BTCV data. Three representative cases are shown. The region with visual improvement is boxed and enlarged. White arrows emphasized the segmentation improvement on portal vein (yellow), stomach (purple), gallbladder (dark green), left kidney (light green), and spleen (red).}
\label{fig:btcv_vis}
\end{figure*}
\subsection{Multi-organ Segmentation}
We present the quantitative performance and qualitative segmentation comparison on the BTCV dataset in Table~\ref{tab:btcv} and Fig.~\ref{fig:btcv_vis}, respectively. No pre-training or ensemble is performed in all experiments. UNesT achieves the best average performance on BTCV dataset which demonstrate the generalizability of UNesT. Compared with the other methods, UNesT achieves large improvement on organs that are small in size, such as the esophagus, pancreas, and adrenal glands, where UNesT outperforms the second best performing method by 2.5\%, 1.2\% and 1.9\%, respectively. In Fig.~\ref{fig:btcv_vis} rows 1 and 2, UNesT successfully differentiates stomach tissues and background tissues demonstrating that UNesT has a better capability on identifying heterogeneous organs. UNesT better captures spatial information in Fig.~\ref{fig:btcv_vis} row 3, where most of the other model confuses right/left kidneys and liver/spleen tissues. 

\begin{table*}[ht]
\caption{Quantitative comparison of the segmentation results on the BTCV testing set. All results shown are single model performances (without ensemble). Our model achieves the overall best performance. Note: RKid: right kidney, LKid: left kidney, Gall: gallbladder, Eso: esophagus, Stom: Stomach, Panc: pancreas, RAG: right adrenal gland, LAG: left adrenal gland.}
\begin{adjustbox}{width=\textwidth}

\label{tab:btcv}
\begin{tabular}{l|rrrrrrrrrrrrr|r}
\hline
Methods   & \multicolumn{1}{l}{Spleen}          & \multicolumn{1}{l}{RKid}   & \multicolumn{1}{l}{LKid}   & \multicolumn{1}{l}{Gall}            & \multicolumn{1}{l}{Eso}    & \multicolumn{1}{l}{Liver}  & \multicolumn{1}{l}{Stom}   & \multicolumn{1}{l}{Aorta}  & \multicolumn{1}{l}{IVC}    & \multicolumn{1}{l}{PSV}    & \multicolumn{1}{l}{Panc}   & \multicolumn{1}{l}{RAG}    & \multicolumn{1}{l|}{LAG}    & \multicolumn{1}{l}{Avg}    \\ \hline
nnUNet    & \multicolumn{1}{l}{\textbf{0.9595}} & \multicolumn{1}{l}{0.8835} & \multicolumn{1}{l}{0.9302} & \multicolumn{1}{l}{\textbf{0.7013}} & \multicolumn{1}{l}{0.7672} & \multicolumn{1}{l}{0.9651} & \multicolumn{1}{l}{0.8679} & \multicolumn{1}{l}{0.8893} & \multicolumn{1}{l}{0.8289} & \multicolumn{1}{l}{0.7851} & \multicolumn{1}{l}{0.7960} & \multicolumn{1}{l}{0.7326} & \multicolumn{1}{l|}{0.6835} & \multicolumn{1}{l}{0.8316} \\
TransBTS  & 0.9455                              & 0.8920                     & 0.9097                     & 0.6838                              & 0.7561                     & 0.9644                     & 0.8352                     & 0.8855                     & 0.8248                     & 0.7421                     & 0.7602                     & 0.6723                     & 0.6703                      & 0.8131                     \\
nnFormer  & 0.9458                              & 0.8862                     & 0.9368                     & 0.6529                              & 0.7622                     & 0.9617                     & 0.8359                     & 0.8909                  & 0.8080                     & 0.7597                     & 0.7787                     & 0.7020                     & 0.6605                      & 0.8162                     \\
CoTr &0.9536 &0.8940 &0.9330 &0.6954 &0.7749 &0.9617 &0.8801 &\textbf{0.9047} &0.8376 &\textbf{0.7891} &0.7964 &0.7350 &0.6831 &0.8356 \\
UNETR     & 0.9048                              & 0.8251                     & 0.8605                     & 0.5823                              & 0.7121                     & 0.9464                     & 0.7206                     & 0.8657                     & 0.7651                     & 0.7037                     & 0.6606                     & 0.6625                     & 0.6304                      & 0.7600                     \\
SwinUNETR & 0.9459                              & 0.8897                     & 0.9239                     & 0.6537                              & 0.7543                     & 0.9561                     & 0.7557                     & 0.8828                     & 0.8161                     & 0.7630                     & 0.7452                     & 0.6823                     & 0.6602                      & 0.8044                     \\ \hline
UNesT     & 0.9580                              & \textbf{0.9249}            & \textbf{0.9396}            & 0.7002                              & \textbf{0.7940}            & \textbf{0.9657}            & \textbf{0.8861}            & 0.8899            & \textbf{0.8412}            & 0.7856            & \textbf{0.8058}            & \textbf{0.7372}            & \textbf{0.7083}             & \textbf{0.8433}            \\ \hline
\end{tabular}
\end{adjustbox}
\end{table*}

\subsection{Kidney and Tumor Segmentation}
\begin{figure}
\centering
\includegraphics[width=0.45\textwidth]{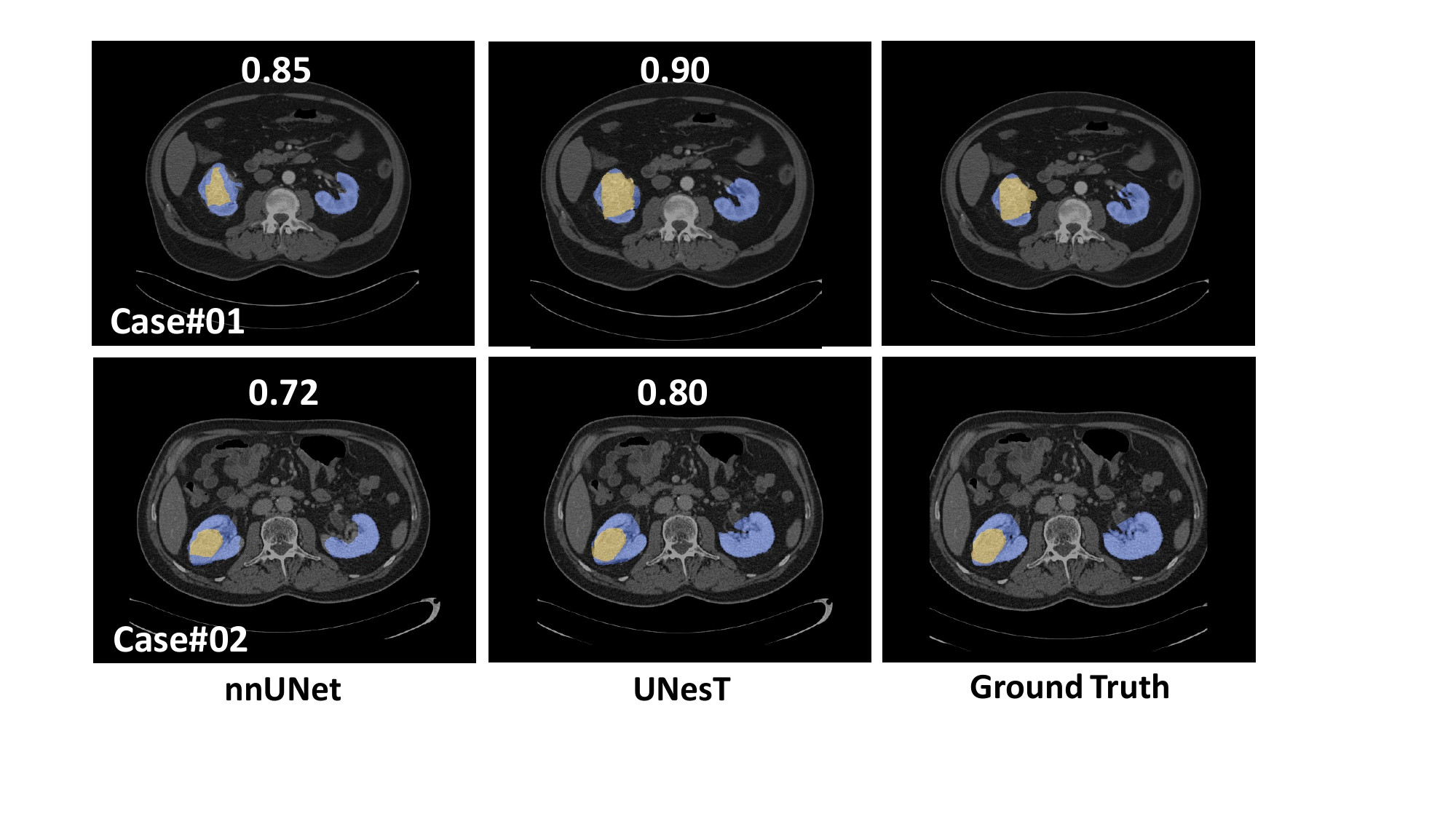}
\caption{Qualitative comparison between our transformer-based segmentation method and the CNN-based model. UNesT shows better tumor segmentation, and we observe the model can better distinguish the kidney-tumor boundary. }
\label{fig:kits_quali}
\end{figure}

To validate the generalizability of UNesT, we compare KiTS19 results among nnUNet~\cite{isensee2021nnu} and transformer-based methods. Our approach achieves moderate improvement at DSC of $0.9794$ and $0.8439$ for kidneys and tumors, respectively, as shown in Table~\ref{tab:kits_quanti}, indicating that the designed architecture can be used as a generic 3D segmentation method. We show a qualitative comparison between our transformer-based model with the CNN-based nnUNeT in Fig.~\ref{fig:kits_quali}. Case $1$ is an above average sample that shows UNesT achieves a clearer boundary between kidney and tumor, while case $2$ is an under average case where the 3D DSC score of UNesT achieves $0.80$ compared to $0.72$. 

\subsection{Brain Tumor Segmentation}
In Table.~{\ref{tab:brats_compare}}, we compared the performance of UNesT with three top-performed methods in BraTS 2021 challenge dataset. Dice scores of the three types of brain tumors are presented in the table with 5 folds experiment design. UNesT consistently outperforms the CNN-based method SegResNet and nnUNet, and the transformer-based Swin UNETR. In particular, for ET, UNesT achieves top Dice scores of 0.898 on average and outperforms the closest competing method by $0.7\%$. Overall, the average Dice across 5 folds and 3 brain tumor structures is 0.917 which surpasses the state-of-the-art by $0.4\%$.
\begin{table}[t]
\centering
\caption{KiTS19 DSC performance comparison with baseline methods. The UNesT achieves the highest DSC on the with-held test set.}
\label{tab:kits_quanti}
\resizebox{\linewidth}{!}{%
\begin{tabular}{c|cc|c}
\hline
Model   & Kidney & Tumor    & Avg              \\ \hline
nnUNeT~\cite{isensee2021nnu} & 0.9643  & 0.8287 & 0.8965     \\
nnFormer~\cite{zhou2021nnformer} & 0.9723  & 0.8348  & 0.9036   \\
CoTr~\cite{xie2021cotr} & 0.9735 &0.8341 &0.9038\\
TransBTS~\cite{wang2021transbts} & 0.9740 & 0.8374 & 0.9057  \\
UNETR~\cite{hatamizadeh2022unetr} & 0.9746 & 0.8382 & 0.9064  \\
Swin UNETR~\cite{tang2022self} & 0.9751 & 0.8397 & 0.9074  \\
\hline 
UNEST (Ours) & \textbf{0.9794} & \textbf{0.8439} & \textbf{0.9117}
\\\hline
\end{tabular}
}
\end{table}

\begin{table*}[]
\centering
\caption{Five-fold cross-validation performance for BraTS 2021 challenge dataset, metrics are Dice scores. Baseline methods benchmarks are directly from respective models trained by challenge participants. Brain tumor regions: ET, WT, and TC denote Enhancing Tumor, Whole Tumor, and Tumor Core.}
\label{tab:brats_compare}
\resizebox{0.99\linewidth}{!}{%
\begin{tabular}{
l |
c
c 
c 
c 
c |
c 
c 
c
c 
c |
c 
c 
c
c 
c |
c
}
\hline
 {}  & \multicolumn{16}{c}{BraTS Challenge 2021} \\ \cline{2-17}
\multirow{-2}{*}{Method}     & \multicolumn{5}{c|}{ET} & \multicolumn{5}{c|}{WT} & \multicolumn{5}{c|}{TC}  & {Avg.}\\ \cline{2-16}
         & 0 & 1 & 2 & 3 & 4  & 0 & 1 & 2 & 3 & 4  & 0 & 1 & 2 & 3 & 4 &  \\ \hline
SwinUNETR~\cite{tang2022self}         & 0.876 & 0.908 & 0.891 & 0.890 & 0.891   & 0.929 & \textbf{0.938}  & 0.931 & \textbf{0.937} & 0.934              & 0.914  & 0.919  & \textbf{0.919} & 0.920 & 0.917 & 0.913 \\
SegResNet~\cite{myronenko20193d}   & 0.867 & 0.900  & 0.884 & 0.888 & 0.878     & 0.924  & 0.933  & 0.927 & 0.921 & 0.930    & 0.907 & 0.915  & 0.917 & 0.916  & 0.912   & 0.907    \\
nnUNET~\cite{isensee2021nnu}           & 0.866 & 0.899 & 0.886 & 0.886  & 0.880     & 0.921 & 0.933  & 0.929  & 0.927 & 0.929   & 0.902 & 0.919 & 0.914  & 0.914  & 0.917 & 0.908    \\
UNesT        & \textbf{0.887} & \textbf{0.915} & \textbf{0.896} & \textbf{0.894}   & \textbf{0.896} & \textbf{0.930} & 0.936  & \textbf{0.939} & 0.924 & \textbf{0.937} & \textbf{0.915} & \textbf{0.923} & \textbf{0.919} & \textbf{0.928} & \textbf{0.923}  & \textbf{0.917}  \\ \hline
\end{tabular}%
}
\end{table*}

\begin{figure*}[h!]
\centering
\includegraphics[width=\textwidth]{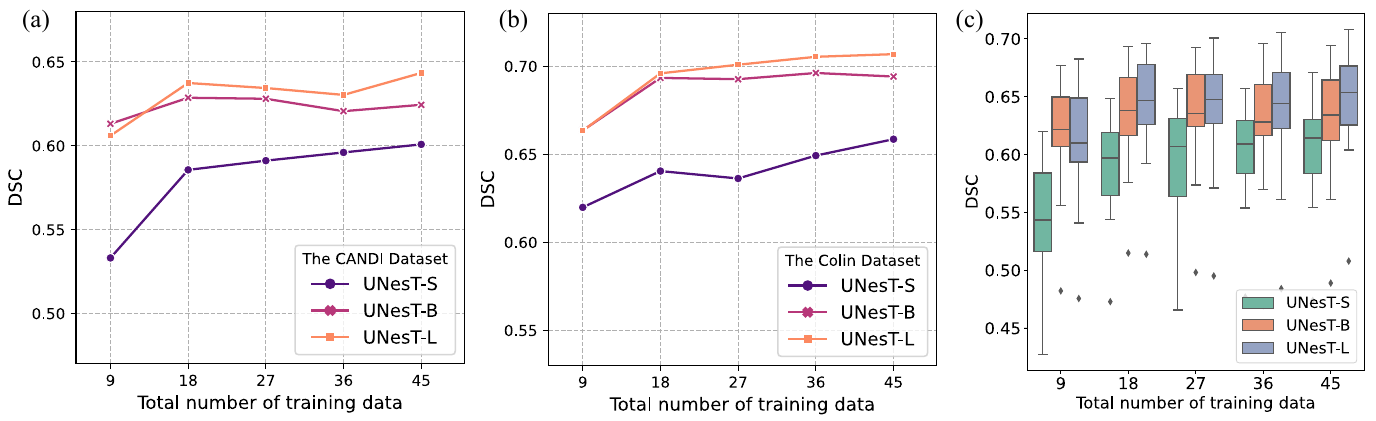}
\caption{Comparison of segmentation results of models with different scales trained with different percentages of training data. (a) and (b) shows the test results of the CANDI and Colin dataset, respectively. (c) shows the results in both the CANDI and Colin dataset. }
\label{fig:modelscale_lineplot}
\end{figure*}

\begin{figure*}[h!]
\centering
\includegraphics[width=0.8\textwidth]{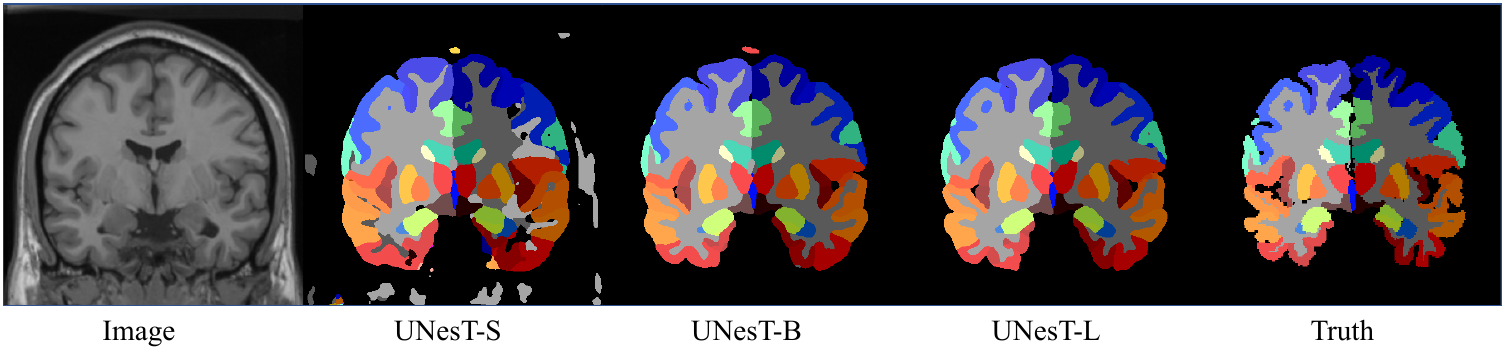}
\caption{Visualization of segmentation results for each model scale trained with the same number of data. Comparing UNesT-S and UNesT-B, UNesT-S evidently mis-classified a large amount of background and brain tissues pixels whereas UNesT-B has mostly clean background indicating that UNesT-B has better capability on utilizing the training data efficiently. UNesT-L further improves the segmentation results indicating that larger models are more data efficient. }
\label{modelscale_vis}
\end{figure*}
\subsection{Ablation Study}

\subsubsection{Model Scales}
To investigate the scalability of our proposed model, we designed "small", "base" and "large" UNesT models (UNesT-S, UNesT-B and UNesT-L) by scaling the depth, heads and width of the transformer. Detailed parameters of UNesT models with various hyperparameter settings are shown in Table~\ref{tab:table1}. Experiments are performed on whole brain segmentation task with 50 T1w MRI scans from the OASIS dataset. 45 T1w scans are used for training and the other 5 for validation. No pre-training is performed for all the models. We start with $20\%$ of the training data and add $20\%$ each time until all data are included. All models are trained five times with 9, 18, 27, 36, and 45 samples, respectively. Fig.~\ref{fig:modelscale_lineplot}(a) and (b) shows the quantitative results of DSC in the CANDI and Colin dataset, respectively. Fig.~\ref{fig:modelscale_lineplot}(c) shows the distribution of the average DSC in each subject of the test set. Fig.~\ref{modelscale_vis} shows the qualitative comparison of whole brain segmentation of different model scales trained with 45 T1w scans.

We observe that larger models and additional data improve segmentation performance. Larger models are more data efficient as with the amount of training data increase, larger models perform better than smaller models. In Fig.~\ref{modelscale_vis}, compared with UNesT-S and UNesT-B, UNesT-S evidently mis-classified a large amount of background and brain tissues pixels whereas UNesT-B has mostly clean background indicating that UNesT-B better utilizes the training data efficiently. UNesT-L further improves the segmentation results indicating that larger models are more data efficient. In terms of reducing annotation effort, both UNesT-B and UNesT-L perform better with 9 training samples than UNesT-S with all the training samples, which reduces the annotation effort by at least 80\%. When adding additional data, the DSC score increases for all the models of different scales. 
In terms of the relationship between model size and DSC score performance, although the DSC score performance steadily increases as the model scale increases, the performance differences become smaller.
In low-data regime, UNesT-B can achieve comparable DSC compared to UNesT-L, but UNesT-B marginally outperforms UNesT-S. When all the training data are included,  the performance increase ratio between UNesT-B and UNesT-S is 5.41\% (0.6941 versus 0.6585) compared with 1.83\% between UNesT-S and UNesT-B (0.7068 versus 0.6941) on the Colin dataset and 3.93\% (0.6244 versus 0.6008) versus 3.04\% (0.6434 versus 0.6244) on the CANDI dataset. Although UNesT-L has 3 times more parameters than UNesT-B, the comparable performance between UNesT-L and UNesT-B indicates UNesT-B is efficient for the training data. After reaching a certain point, scaling up models may not necessarily lead to large performance improvements.

In Table.~{\ref{tab:time}}, we summarize the training and testing/inference time with the whole brain segmentation using different model scales and sliding window overlaps. According to our benchmarks, the inference time is mostly impacted by the sliding window overlap, as the increase of overlap will result in a significant number of patches in $overlap=0.7$. We benchmark the registered MRI volume $172 \times 220 \times 156$. In practical clinical settings, auto segmentation of a given MRI volume with a single GPU or CPU can achieve satisfactory performance (shown in the Fig.~{\ref{fig:modelscale_lineplot}}) and inference time (e.g., 2.34s).

\begin{table}[t]
\centering
\caption{Model parameters of different scales. Depth: number of transformer layers, Heads: number of heads in the multi-head attention, Width: embedded dimension. Each number in the bracket represents the corresponding hyperparameter in that hierarchy.}
\label{tab:table1}
\begin{tabular}{ccccc}
\hline
Model   & \#Param & Depth    & Heads     & Width         \\ \hline
UNesT-S & 22.4M  & (2,2,8)  & (2,4,8)   & (64,128,256)  \\
UNesT-B & 87.3M  & (2,2,8)  & (4,8,16)  & (128,256,512) \\
UNesT-L & 279.6M & (2,2,20) & (6,12,24) & (192,384,768) \\\hline 
\end{tabular}
\end{table}

\begin{table}[t]
\centering
\caption{Training and inference time of different model scales. Training iteration time reports the time for processing a single $96 \times 96 \times 96$ patch. For the testing cases, the latencies are reported on MNI space size of $172 \times 220 \times 156$ using GPU RTX 3090Ti, regular inference pipeline without mixed precision, and no torchscript and TensorRT conversions. Note the inference patch size is $96 \times 96 \times 96$, sliding window overlap has a significant impact on the inference time because the increase of overlap percentage will result in exponentially increased patches. }
\label{tab:time}
\begin{tabular}{
l 
l 
l 
l }
\hline
Model                                                      & UNesT-S & UNesT-B & UNesT-L \\ \hline
\multicolumn{4}{c}{Training}                                     \\ \hline
\multicolumn{1}{l|}{iteration(s)}  & 0.29    & 0.46    & 0.82    \\ \hline
\multicolumn{4}{c}{Testing}                                      \\ \hline
\multicolumn{1}{l|}{overlap=0.3(s)}   & 0.84       & 0.98       & 1.23       \\
\multicolumn{1}{l|}{overlap=0.5(s)}   & 2.30       & 2.34       & 2.97       \\
\multicolumn{1}{l|}{overlap=0.7(s)}   & 6.76       & 6.99       & 7.85       \\
 \hline
\end{tabular}
\end{table}

\subsubsection {Data Efficiency}
\begin{figure*}[h!]
\centering
\includegraphics[width=\textwidth]{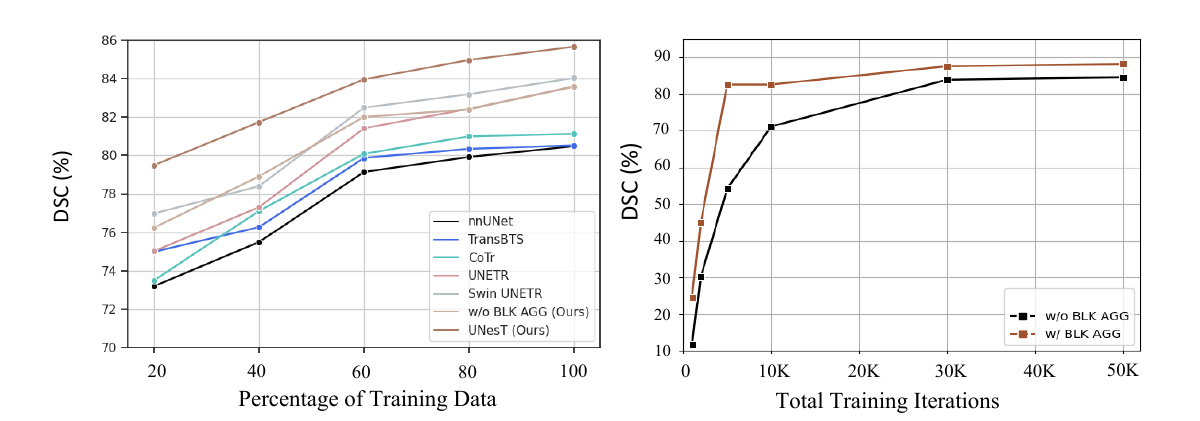}
\caption{Left: DSC comparison on the test set at different percentages of training samples. Right: Comparison of the convergence rate for the proposed method with and without hierarchical modules, and validation DSC along training iterations are demonstrated. Different ViT-, CNN-based and hybrid baselines are compared.}

\label{fig:fig4}
\end{figure*}

We investigate the data efficiency of our proposed method using the whole brain and renal substructures dataset. Fig.~\ref{modelscale_vis} shows the performance comparison between different UNesT variants, base and larger model are of better data-efficient when training with less data (e.g., 9 or 18). We show the \underline{UNesT-B model achieves 133 classes segmentation of DSC} \underline{0.6131 with only 9 training samples.} 
Figure~\ref{fig:fig4} shows the data efficiency evaluated and compared on the renal substructure dataset. UNesT achieves DSC of 0.7903 compared to the second-best SwinUNETR 0.7681 when training with 20\% samples. With the increase of training data, our method performs consistently higher DSC compare to baseline methods. We observe the UNesT model trained with 20\% data is comparable to nnUNet or TransBTS using full training data, which shows superior data efficiency. 

\subsubsection {Effects of Block Aggregation}
We show the hierarchical architecture design (with 3D block aggregation) provides significant improvement for medical image segmentation (as shown in Fig.~\ref{fig:fig4}). The result shows that the hierarchy mechanism achieves superior performance at 20\% to 100\% of training data. Under a low-data regime, block aggregation achieves a higher improvement ($> 3\%$ of DSC) compared to the second-best method. We notice that the model without block aggregation (canonical transformer layers) obtains lower performance. In addition, UNesT with block aggregation demonstrates a faster convergence rate (15\% and 4\% difference at 2K/30K iterations) compared to the backbone model without hierarchies. The results show block aggregation is an effective component for representation learning for transformer-based models. In addition, compared with the Swintransformer-based method, our UNesT shows consistently superior performance, especially in whole brain segmentation, which indicates the 3D aggregation modules perform better than shifted window module for local patch communication. 

\subsubsection {Size of pre-training dataset in whole brain segmentation}
\begin{figure}
\centering
\includegraphics[width=0.45\textwidth]{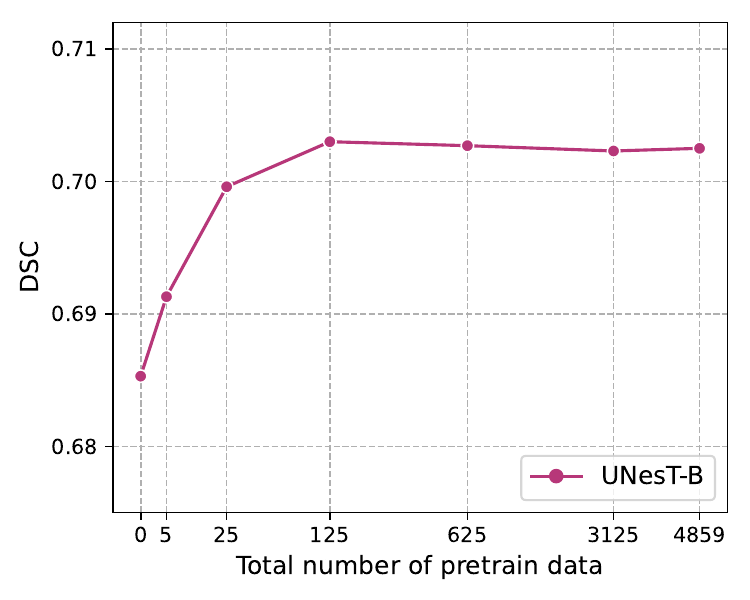}
\caption{DSC comparison on whole brain segmentation CANDI dataset with different amount of data with pseudo labels for pre-training. }
\label{fig:pretrain}
\end{figure}

Acquiring human annotation is labor intensive, thus many studies \citep{yang2022label,roy2017error,huo20193d,yang2022quantification} adopt the strategy of pre-training with pseudo labels and then finetuning with human annotations to get around this limitation and increase model performance. Herein, we perform experiments using different amounts of pre-training data in the whole brain segmentation to investigate the impact of pre-training data quantity on the final results. We repeat the experiments 6 times with 5, 25, 125, 625, 3125, and all available pre-training data, respectively. All the pre-trained models are finetuned using the OASIS dataset in the same 5-fold cross-validation setting. We include the results of the training from scratch using the OASIS dataset for comparison. The results are shown in Fig.~\ref{fig:pretrain}. We observe that increasing the number of pre-trained examples up to 125 resulted in a rapid improvement in the DSC score. Pre-training sizes greater than 125 do not further advance performance and the results fluctuate in a small range. This observation demonstrates that UNesT can benefit from pre-training using pseudo data, but a large pre-training dataset is not a necessity. When the amount of pre-training data reaches a certain limit, the performance gains are reduced. Instead, adding more pseudo data could possibly confuse the network.

\section{Discussion}
\subsection{Why do we need an efficient hierarchical transformer-based medical segmentation model?}
In this paper, we target the critical problem that transformed-based models commonly lack of local positional information resulting in sub-optimal performance when handling considerable tissue classes in 3D medical image segmentation. Specifically, medical segmentation datasets are small where images are of spatially high-resolution and high dimensionality which can lead to data inefficiency. Our proposed UNesT addresses the above problem by hierarchically aggregating the spatially adjacent patches and leveraging the global self-attention mechanism to combine global and local information efficiently. SwinUNETR~\cite{tang2022self}, which uses "shifted window" for local patch communication, observed good but inconsistent performance. Specifically, it achieves second-best performance in the renal substructures segmentation, but in the whole brain segmentation, its DSC scores in test datasets under-perform the second-best performing SLANT27 model by a large margin. Our method consistently achieves superior results on the four evaluated heterogeneous tasks.


We highlight our method on dealing with multiple tissues and inter-connected structures. Compared to the prior state-of-the-art method SLANT27~\cite{huo20193d}, which used 27 ensembled networks, UNesT successfully achieves better performance with a single model. Among current 3D medical image segmentation methods, we address the challenging tasks, including more than one hundred structures in T1w MRI, three inter-connected components in kidneys, thirteen major organs in the abdomen, kidney-tumor, and brain-tumor connected tissue.

When predicting multiple tissues that have various sizes and shapes simultaneously, the model performance is often susceptible to the tissues that are small in size. And when dealing with inter-connected tissues, model performance is particularly sensitive to boundary prediction, where missing boundary prediction will jeopardize the results of adjacent classes. However, boundary prediction is not a trivial task in medical image segmentation since the images usually have blurry boundaries and similar intensity/appearance which make it hard to characterize one tissue from another. In our UNesT model, we adopt hierarchical design which utilizes multi-scale strategy to handle the difference in tissue size and blurry boundary problems. At a coarse scale, the model can focus more on the overall structure of the image, and at a finer scale, the model can focus more on the detail of the tissues. Additionally, our design of 3D block aggregation provides additional adjacent positional information to the model, which gives it better tissue distinguish capability. Therefore, UNesT achieves better performance on handling multiple tissues and inter-connected tissue problems.

UNesT shows consistent competitive performance for the brain tumor segmentation task, which is a difficult problem. UNesT contains several hierarchical blocks as its encoder, and it can efficiently encode the multi-scale features of the 3D multi-modal inputs. And multi-scale embeddings are of significant value to medical image segmentation. We also observe that Swin UNETR, SegResNet, and nnU-Net achieve close competitive performance in this dataset. The three baselines contain feature downsample and upsample modules, where multi-scale feature maps are utilized and output to the decoder. The quantitative results show that our method can be effective at modeling tumor tissues and efficiently learning multi-scale features.

The superior performance of UNesT in these inferences 5 different tasks demonstrates the effectiveness and efficiency of UNesT in segmenting multiple structures/tissues with small medical datasets. Specifically, we validate that our model is data efficient in low-data regimes. Moreover, our experiments show that larger models are more data efficient, suggesting the proposed network is easily scalable if necessary. Furthermore, we study the impact of the number of pseudo labels used for pre-training. We observed that pre-training sizes exceeding a certain number do not further advance model performance. On the contrary, adding more pseudo labels may confuse the network and decrease performance (Fig.~\ref{fig:pretrain}). 

\subsection{The Combination of CNN and Transformers}

The proposed UNesT follows the first class of the CNNs and transformer combination design where transformer is the main encoder and CNNs served as the decoder. To evaluate the effectiveness of our proposed UNesT in comparison to other CNN and transformer hybrid models, we compare two models falling under the first category - UNETR and SwinUNETR. Additionally, we include TransBTS and CoTr from the second category, where transformers serve as secondary encoders and CNNs serve as the main encoder and decoder. All these baseline methods are specifically designed for 3D volumetric medical image segmentation. As there are currently no existing techniques in the third category that are optimized for this task, we do not include them in our comparative analysis. 

CoTr achieves good performance in whole brain segmentation and multi-organ segmentation task in terms of DSC. However, in whole brain segmentation, both TransBTS and CoTr have inf value in the HD of the CANDI dataset. This may indicate that this type of model is susceptible to outliers. On the other hand, UNETR, SwinUNETR, and UNesT achieve relatively stable performance among four datasets, especially when dealing with outlier cases in the whole brain segmentation task. 

\subsection{Single Models Performance vs. Ensembles}
According to the single instance benchmarks (Table.~{\ref{tab:whole_single}} and Table.~{\ref{tab:kits_single}}), we observe a similar performance trend to what is observed in the ensemble results (Table.~{\ref{tab:brainresults}}, Table.~{\ref{tab:tab1}}) and Table.~{\ref{tab:kits_quanti}}. Some models have improved performance without using an ensemble, while others experience a minor decrease in performance. In general, the model's performance is comparable whether or not an ensemble is utilized, but the use of an ensemble could potentially enhance the stability of the model's performance. Ensemble five-fold models can effectively remove outlier predictions and improve overall performance. With single model testing, our method consistently achieves the best performance across different datasets. Compared to the five-fold ensemble strategy, a single instance is more practical and used in clinical workflow with faster inference time and less computational effort.

\begin{table}[]
\centering
\caption{Single model performance for the whole brain segmentation task. Models are trained with the same training/validation data.}
\label{tab:whole_single}
\resizebox{0.99\linewidth}{!}{%
\begin{tabular}{
l |
c 
c 
c 
c }
\hline
                                & \multicolumn{2}{c}{Colin} & \multicolumn{2}{c}{CANDI} \\ \cline{2-5} 
\multirow{-2}{*}{Method}        & DSC                     & HD                      & DSC                     & HD                      \\ \hline
nnUNet~\cite{isensee2021nnu}    & 0.7062                  & 14.0101                 & 0.3930                  & inf                     \\
TransBTS~\cite{wang2021transbts}  & 0.6542                  & inf                     & 0.5991                  & inf                     \\
nnFormer~\cite{zhou2021nnformer}  & 0.7007                  & 10.423                  & 0.6420                  & inf                     \\
CoTr~\cite{xie2021cotr}          & 0.7268                  & 10.2561                 & 0.6923                  & inf                     \\
UNETR~\cite{hatamizadeh2022unetr} & 0.7328                  & 10.216                  & 0.6810                  & 13.3172                 \\
SwinUNETR~\cite{tang2022self}    & 0.6853                  & 21.4812                 & 0.6536                  & 34.5212                 \\
SLANT~\cite{huo20193d}                                                  & 0.7301                  & \textbf{9.9470}         & 0.6977                  & 9.5000                  \\\hline
UNesT                                                   & \textbf{0.7467}         & 11.0358                 & \textbf{0.7022}         & \textbf{8.8902}         \\ \hline
\end{tabular}%
}
\end{table}

\begin{table*}[]
\centering
\caption{Single model performance for the KiTS19 task and renal substructure segmentation. Models are trained with the same training/validation data. Pel. Sys. refers to Pelvicalyceal System.}
\label{tab:kits_single}
\resizebox{0.8\linewidth}{!}{%
\begin{tabular}{
l |
c 
c 
c |
c 
c 
c 
c
}
\hline
 {}    &  \multicolumn{3}{c|}{KiTS19} & \multicolumn{4}{c}{Renal Substructures} \\ \cline{2-8}
\multirow{-2}{*}{Method}     & {Kidney} & {Tumor} & {Avg.} & {Cortex} & {Medulla} & {Pel. Sys.} & {Avg.}\\ \hline 
nnUNet~\cite{isensee2021nnu}    & 0.9640                  & 0.8198                               & 0.8919          & 0.8881 & 0.7974 & 0.7285 & 0.8047    \\
nnFormer~\cite{zhou2021nnformer}  & 0.9714                  & 0.8321                                    & 0.9018    & 0.9082 & 0.8076 & 0.7370 &  0.8176        \\
CoTr~\cite{xie2021cotr}   & 0.9706                  & 0.8336                                & 0.9021             & 0.8950 & 0.7987 & 0.7304 & 0.8080   \\
TransBTS~\cite{wang2021transbts}       & 0.9710                  & 0.8358                               &  0.9034        & 0.8884 & 0.8009 & 0.7271 & 0.8055    \\
UNETR~\cite{hatamizadeh2022unetr} & 0.9737                  & 0.8362                             & 0.9050      & 0.9015 & 0.8143 & 0.7577 & 0.8245        \\
SwinUNETR~\cite{tang2022self}    & 0.9739                  & 0.8368                              & 0.9054       & 0.9040 & 0.8302 & 0.7603 & 0.8315    \\ \hline
UNesT                                                   & \textbf{0.9790}                      & \textbf{0.8434}         & \textbf{0.9112}     & \textbf{0.9211} & \textbf{0.8399} & \textbf{0.7906} & \textbf{0.8505}    \\ \hline
\end{tabular}%
}
\end{table*}
\subsection{Reproducibility against clinical radiologists} In this work, we develop the first in-house renal sub-structures CT cohort for segmentation, including the renal cortex, medulla, and pelvicalyceal system which are manually annotated by radiologists. We show that the proposed method is data-efficient for accurately quantifying kidney components and can be used for volumetric analysis such as in the medullary pyramids. Fig.~\ref{fig:figA1} shows the proposed automatic segmentation method achieves better agreement compared to inter-rater assessment, with 0.03 versus 0.29 mean difference, respectively, indicating robust reproducibility. Visual quantitative analysis of renal structures remains a complex task for radiologists. Some of the histomorphometric features in regions of the kidney (e.g., textural or graph features) are poorly adapted for manual identification. In this study, we show that UNesT achieves consistently reliable performance. Compared with previous studies on cortex segmentation, the proposed approach significantly facilitates the derivation of the visual and quantitative results.

\subsection{Limitation and sensitivity study} 
For whole brain segmentation, we observe current performance is limited by registration. Specifically, the DSC score in the MNI space is around 0.90 and around 0.87 in the Colin and CANDI dataset, respectively. However, the performance drops around 0.17 DSC score after inverse transformation to the original space. Investigation of registration performance should be considered in the future.

\begin{figure}[htp]
\centering
\includegraphics[width=\linewidth]{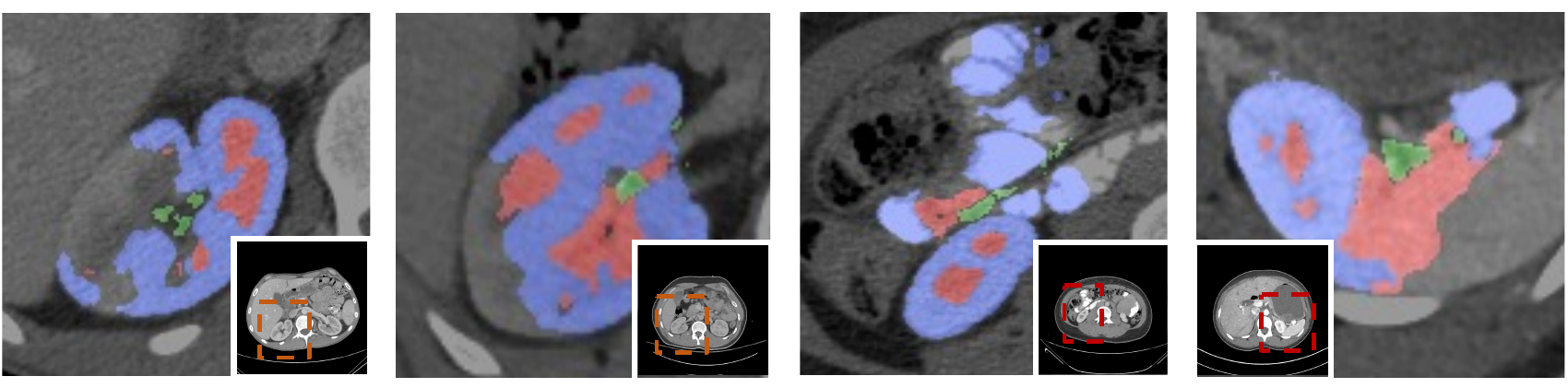}
\caption{Demonstration of potential outlier cases. The left two images show representative under-labeling of tissues. The right two images show the over-labeling of tissues. These segmentations are computed on additional contrast-enhanced CT scans without ground truth labels. }
\label{fig:limitation}
\end{figure}

We study outlier cases of renal structure segmentation to demonstrate potential limitations. In reviewing most computer-automated segmentation methods, we found about $90\%$ of the segmentation is promising, but about $10\%$ are also found to be outliers. As shown in Fig.~\ref{fig:limitation}, typical outliers under-segment and fail to capture parts of tissue labels (left two images). The missing parts result in a lower DSC score of about 0.80 (cortex) and 0.62 (medulla). The right two images show the other type of failure: over-segmentation, where we observe a complete renal segmentation but mis-labeling of nearby tissues. This issue can potentially be resolved by component analysis in a post-processing step. These two types of outlier segmentation are easily spotted with a rudimentary visual quality check. 

\section{Conclusions}
In this paper, we propose a novel hierarchical transformer-based 3D medical image segmentation approach (UNesT) with a 3D block aggregation module to achieve local communication. We validate the effectiveness of UNesT on 5 different tasks in both CT and MRI modalities including a whole brain segmentation task with 133 classes, a renal substructure segmentation task, a multi-organ abdominal segmentation task, and a kidney/tumor segmentation task as well as a brain tumor segmentation task. We consistently achieve state-of-the-art performance on the four datasets. Our single model outperforms $27$ ensembled models in the prior state-of-the-art method, SLANT27, for whole brain segmentation. In addition, we develop the first in-house renal sub-structures CT dataset with radiologists. UNesT achieves the best performance among recent popular convolutional- and transformer-based volumetric medical segmentation methods. We show the major contribution of the proposed method on successfully modeling hundreds of tissues (e.g., 133 classes) and hierarchically inter-connected structures.

\section*{Acknowledgments}
This research is supported by NIH Common Fund and National Institute of Diabetes, Digestive and Kidney Diseases U54DK120058, NSF CAREER 1452485, NIH grants, 2R01EB006136, 
1R01EB017230 (Landman), and R01NS09529. The identified datasets used for the analysis described were obtained from the Research Derivative (RD), database of clinical and related data. The imaging dataset(s) used for the analysis described were obtained from ImageVU, a research repository of medical imaging data and image-related metadata. ImageVU and RD are supported by the VICTR CTSA award (ULTR000445 from NCATS/NIH) and Vanderbilt University Medical Center institutional funding. ImageVU pilot work was also funded by PCORI (contract CDRN-1306-04869).





\bibliographystyle{model2-names.bst}\biboptions{authoryear}
\bibliography{refs}

\clearpage
\section*{Appendix}
\appendix
\renewcommand{\thesection}{\Alph{section}}
This work is the first effort to evaluate comprehensive benchmarks for whole brain segmentation and renal substructure segmentation for 3D medical image analysis. We extend the most representative and challenging whole brain segmentation task's performance and show complete performance table for all 133 structures in Appendix~\ref{appendix_a}. In section~\ref{appendix_b}, we introduce the open-source project availability development based on UNesT and the study. 

\section{Whole Brain Segmentation Details}
\label{appendix_a}
\subsection{Data Pre-processing}
We apply N4 bias field correction~\cite{tustison2010n4itk} on the images after registration. We use robust regression (robustfit) in Matlab to normalize the MRI scans in the test set. To train the regression model, inputs are normalized by: $I^{'} = \frac{I - \mu}{\sigma}$, where $I$ is the original image, $I^{'}$ is the normalized output image, $\mu$ is the image mean, and $\sigma$ is the image standard deviation. $I^{'}$ is then masked by a binary map where the brain tissues label probability larger than 0.5 is 1 otherwise 0 to get $I_{mask}$. The sorted $I_{mask}$ is averaged by all the images in the training set to train the regression model. During the testing, the weights are learned gradually. The learned weights are used to normalize test scans. Please refer to \cite{huo20193d} for more details.

\subsection{Detailed Brain Regions and Performance}
A detailed brain regions included in the whole brain segmentation task is shown in Table~\ref{tab:brainlabel}, in which 30, 31, and 32 are not presented in both the Colin and the CANDI dataset and thus excluded for evaluation. Quantitative performance comparison of 130 classes is shown in Fig.~\ref{fig:133boxplot}.

\section{Open-Source Availability}
\label{appendix_b}

For developing publicly available segmentation tools, we introduce the MONAI Bundle module that supports building Python-based workflows via structured configurations. The authors contribute and develop the bundle code that  benefits in four-fold:

\begin{itemize}

\item Run as normal Python command, integrated into MONAI package, no additional installations.
\item Compare to Docker container, Bundle configuration provides good readability and usability by separating system parameter settings from the Python code.
\item The pipeline demonstrates workflow at a higher level and allows for changes and customized implementations.
\item Support labeling with visualization tool plugins such as 3D Slicer.

\end{itemize}

The tutorial and release of code/model for whole brain segmentation using the 3D transformer-based segmentation model UNEST is released at: 

\href{https://github.com/Project-MONAI/model-zoo/tree/dev/models/wholeBrainSeg_Large_UNEST_segmentation}{wholeBrainSeg-Large-UNEST-segmentation}. 

The renal substructure tutorial is released at: 

\href{https://github.com/Project-MONAI/model-zoo/tree/dev/models/renalStructures_UNEST_segmentation}{renalStructures-UNEST-segmentation}

\subsection{Run with python command}
Follow and set path for environments and run command: 

\begin{verbatim}
  python -m monai.bundle run evaluating 
  --meta_file configs/metadata.json 
  --config_file configs/inference.json 
  --loggings_file configs/logging.conf
\end{verbatim}

\subsection{Run with 3D Slicer and MONAI Label}

\begin{figure}[h!]
\centering
\includegraphics[width=0.98\linewidth]{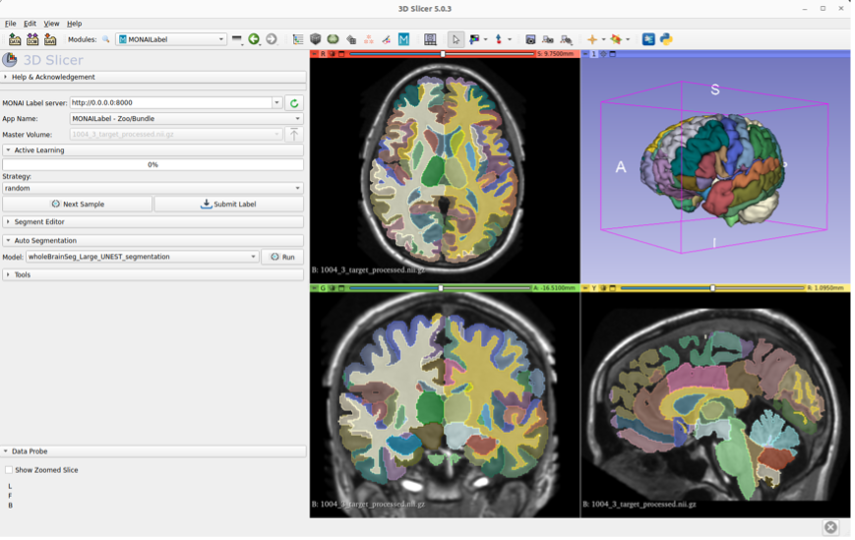}
\caption{Server-based whole brain segmentation using large UNesT annotation model and inference with 3D Slicer.}
\label{fig:appendix_2}
\end{figure}

\begin{figure}[h!]
\centering
\includegraphics[width=0.85\linewidth]{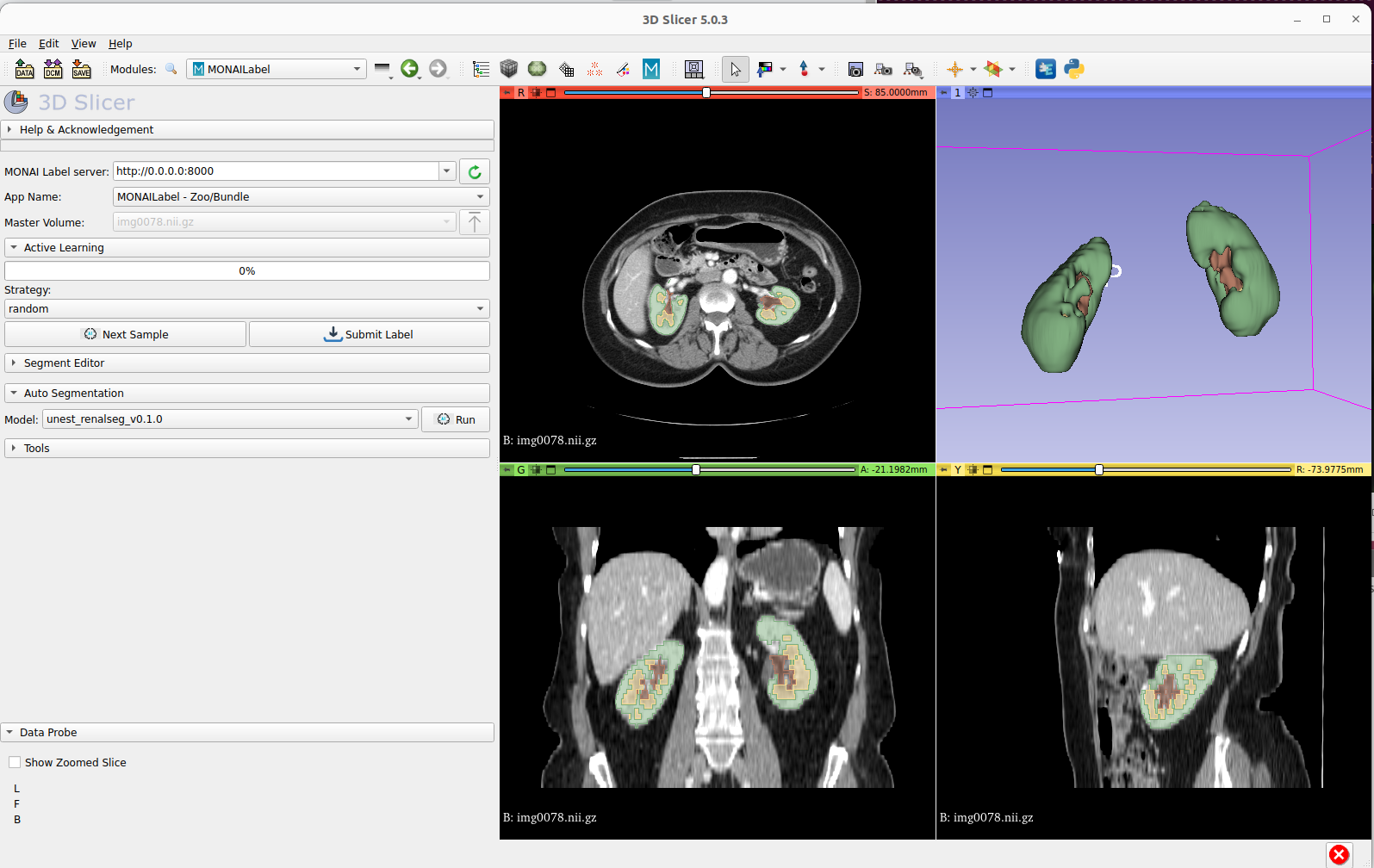}
\caption{Server-based renal substructure segmentation using base UNesT annotation model and inference with 3D Slicer}
\label{fig:appendix_renal}
\end{figure}

\begin{figure*}[h!]
\centering
\renewcommand{\thefigure}{A.\arabic{figure}}
\includegraphics[width=0.68\textwidth]{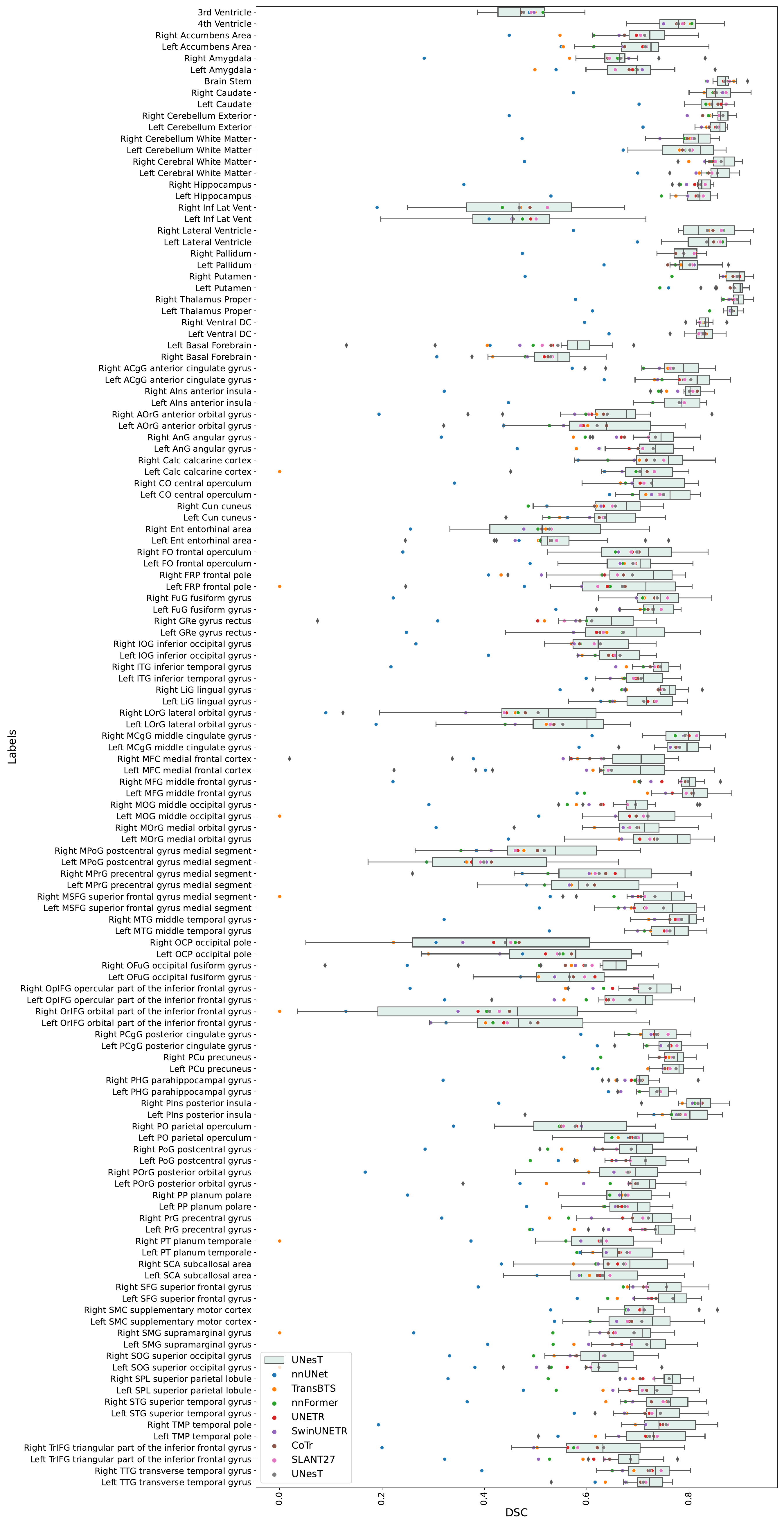}
\caption{Quantitative results of 130 classes in the whole brain segmentation task. Scatters show the mean DSC score of each method. Overall UNesT has the most best-performing classes.}
\label{fig:133boxplot}
\end{figure*}

We developed a server-based tool for integrating annotation whole brain and renal substructures purpose as shown in Figure.~\ref{fig:appendix_2} and~\ref{fig:appendix_renal}. Install MONAI Label to work on labeling with public visualization tools such as 3D Slicer, renal structures segmentation as the \href{https://docs.monai.io/projects/label/en/latest/quickstart.html#tutorial-2-bundle-app-tutorial-and-use-cases}{Tutorial}.
Run command to start the server with renal or whole brain segmentation with UNesT model:
\begin{verbatim}
# start the bundle app in MONAI label

  monailabel start_server --app <full path 
  to the monaibundle app/monaibundle> 
  --studies <path to the local dataset>
  --conf models 
  renalStructures_UNEST_segmentation_v0.1.0
\end{verbatim}


\begin{table*}[ht]
\renewcommand{\thetable}{A.\arabic{table}}  
\caption{List of classes in the whole brain segmentation task. Note: "*" represent the 3 classes not included in the Colin and CANDI dataset.}
\label{tab:brainlabel}
\centering

\begin{adjustbox}{width=0.7\textwidth}
\begin{tabular}{|l|l|l|l|}
\hline
0   & Background                          & 67  & Right MCgG middle cingulate gyrus                         \\ \hline
1   & 3rd Ventricle                       & 68  & Left MCgG middle cingulate gyrus                          \\ \hline
2   & 4th Ventricle                       & 69  & Right MFC medial frontal cortex                           \\ \hline
3   & Right Accumbens Area                & 70  & Left MFC medial frontal cortex                            \\ \hline
4   & Left Accumbens Area                 & 71  & Right MFG middle frontal gyrus                            \\ \hline
5   & Right Amygdala                      & 72  & Left MFG middle frontal gyrus                             \\ \hline
6   & Left Amygdala                       & 73  & Right MOG middle occipital gyrus                          \\ \hline
7   & Brain Stem                          & 74  & Left MOG middle occipital gyrus                           \\ \hline
8   & Right Caudate                       & 75  & Right MOrG medial orbital gyrus                           \\ \hline
9   & Left Caudate                        & 76  & Left MOrG medial orbital gyrus                            \\ \hline
10  & Right Cerebellum Exterior           & 77  & Right MPoG postcentral gyrus medial segment               \\ \hline
11  & Left Cerebellum Exterior            & 78  & Left MPoG postcentral gyrus medial segment                \\ \hline
12  & Right Cerebellum White Matter       & 79  & Right MPrG precentral gyrus medial segment                \\ \hline
13  & Left Cerebellum White Matter        & 80  & Left MPrG precentral gyrus medial segment                 \\ \hline
14  & Right Cerebral White Matter         & 81  & Right MSFG superior frontal gyrus medial segment          \\ \hline
15  & Left Cerebral White Matter          & 82  & Left MSFG superior frontal gyrus medial segment           \\ \hline
16  & Right Hippocampus                   & 83  & Right MTG middle temporal gyrus                           \\ \hline
17  & Left Hippocampus                    & 84  & Left MTG middle temporal gyrus                            \\ \hline
18  & Right Inf Lat Vent                  & 85  & Right OCP occipital pole                                  \\ \hline
19  & Left Inf Lat Vent                   & 86  & Left OCP occipital pole                                   \\ \hline
20  & Right Lateral Ventricle             & 87  & Right OFuG occipital fusiform gyrus                       \\ \hline
21  & Left Lateral Ventricle              & 88  & Left OFuG occipital fusiform gyrus                        \\ \hline
22  & Right Pallidum                      & 89  & Right OpIFG opercular part of the inferior frontal gyrus  \\ \hline
23  & Left Pallidum                       & 90  & Left OpIFG opercular part of the inferior frontal gyrus   \\ \hline
24  & Right Putamen                       & 91  & Right OrIFG orbital part of the inferior frontal gyrus    \\ \hline
25  & Left Putamen                        & 92  & Left OrIFG orbital part of the inferior frontal gyrus     \\ \hline
26  & Right Thalamus Proper               & 93  & Right PCgG posterior cingulate gyrus                      \\ \hline
27  & Left Thalamus Proper                & 94  & Left PCgG posterior cingulate gyrus                       \\ \hline
28  & Right Ventral DC                    & 95  & Right PCu precuneus                                       \\ \hline
29  & Left Ventral DC                     & 96  & Left PCu precuneus                                        \\ \hline
30* & Cerebellar Vermal Lobules I-V       & 97  & Right PHG parahippocampal gyrus                           \\ \hline
31* & Cerebellar Vermal Lobules VI-VII    & 98  & Left PHG parahippocampal gyrus                            \\ \hline
32* & Cerebellar Vermal Lobules VIII-X    & 99  & Right PIns posterior insula                               \\ \hline
33  & Left Basal Forebrain                & 100 & Left PIns posterior insula                                \\ \hline
34  & Right Basal Forebrain               & 101 & Right PO parietal operculum                               \\ \hline
35  & Right ACgG anterior cingulate gyrus & 102 & Left PO parietal operculum                                \\ \hline
36  & Left ACgG anterior cingulate gyrus  & 103 & Right PoG postcentral gyrus                               \\ \hline
37  & Right AIns anterior insula          & 104 & Left PoG postcentral gyrus                                \\ \hline
38  & Left AIns anterior insula           & 105 & Right POrG posterior orbital gyrus                        \\ \hline
39  & Right AOrG anterior orbital gyrus   & 106 & Left POrG posterior orbital gyrus                         \\ \hline
40  & Left AOrG anterior orbital gyrus    & 107 & Right PP planum polare                                    \\ \hline
41  & Right AnG angular gyrus             & 108 & Left PP planum polare                                     \\ \hline
42  & Left AnG angular gyrus              & 109 & Right PrG precentral gyrus                                \\ \hline
43  & Right Calc calcarine cortex         & 110 & Left PrG precentral gyrus                                 \\ \hline
44  & Left Calc calcarine cortex          & 111 & Right PT planum temporale                                 \\ \hline
45  & Right CO central operculum          & 112 & Left PT planum temporale                                  \\ \hline
46  & Left CO central operculum           & 113 & Right SCA subcallosal area                                \\ \hline
47  & Right Cun cuneus                    & 114 & Left SCA subcallosal area                                 \\ \hline
48  & Left Cun cuneus                     & 115 & Right SFG superior frontal gyrus                          \\ \hline
49  & Right Ent entorhinal area           & 116 & Left SFG superior frontal gyrus                           \\ \hline
50  & Left Ent entorhinal area            & 117 & Right SMC supplementary motor cortex                      \\ \hline
51  & Right FO frontal operculum          & 118 & Left SMC supplementary motor cortex                       \\ \hline
52  & Left FO frontal operculum           & 119 & Right SMG supramarginal gyrus                             \\ \hline
53  & Right FRP frontal pole              & 120 & Left SMG supramarginal gyrus                              \\ \hline
54  & Left FRP frontal pole               & 121 & Right SOG superior occipital gyrus                        \\ \hline
55  & Right FuG fusiform gyrus            & 122 & Left SOG superior occipital gyrus                         \\ \hline
56  & Left FuG fusiform gyrus             & 123 & Right SPL superior parietal lobule                        \\ \hline
57  & Right GRe gyrus rectus              & 124 & Left SPL superior parietal lobule                         \\ \hline
58  & Left GRe gyrus rectus               & 125 & Right STG superior temporal gyrus                         \\ \hline
59  & Right IOG inferior occipital gyrus  & 126 & Left STG superior temporal gyrus                          \\ \hline
60  & Left IOG inferior occipital gyrus   & 127 & Right TMP temporal pole                                   \\ \hline
61  & Right ITG inferior temporal gyrus   & 128 & Left TMP temporal pole                                    \\ \hline
62  & Left ITG inferior temporal gyrus    & 129 & Right TrIFG triangular part of the inferior frontal gyrus \\ \hline
63  & Right LiG lingual gyrus             & 130 & Left TrIFG triangular part of the inferior frontal gyrus  \\ \hline
64  & Left LiG lingual gyrus              & 131 & Right TTG transverse temporal gyrus                       \\ \hline
65  & Right LOrG lateral orbital gyrus    & 132 & Left TTG transverse temporal gyrus                        \\ \hline
66  & Left LOrG lateral orbital gyrus     &     &                                                  \\ \hline
\end{tabular}
\end{adjustbox}  
\end{table*}
\end{document}